\documentclass[12pt,a4paper]{article}

\usepackage{amsmath,amssymb}
\usepackage{graphicx}
\usepackage[nosort]{cite}
\usepackage[hyperref,bulletsep]{collect}

\makeatletter
\let\old@startsection=\@startsection
\renewcommand{\@startsection}[6]{\old@startsection{#1}{#2}{#3}{#4}{#5}{#6\mathversion{bold}}}
\makeatother

\makeatletter \@addtoreset{equation}{section} \makeatother

\makeatletter
\let\old@makecaption=\@makecaption
\def\@makecaption{\small\old@makecaption}
\makeatother

\let\oldPhi=\Phi
\let\oldPsi=\Psi
\let\oldGamma=\Gamma
\let\oldDelta=\Delta
\let\oldSigma=\Sigma
\let\oldTheta=\Theta
\let\oldPi=\Pi
\renewcommand{\Phi}{\mathnormal{\oldPhi}}
\renewcommand{\Psi}{\mathnormal{\oldPsi}}
\renewcommand{\Gamma}{\mathnormal{\oldGamma}}
\renewcommand{\Sigma}{\mathnormal{\oldSigma}}
\renewcommand{\Delta}{\mathnormal{\oldDelta}}
\renewcommand{\Theta}{\mathnormal{\oldTheta}}
\renewcommand{\Pi}{\mathnormal{\oldPi}}

\newcommand{\superN}{\mathcal{N}}

\newcommand{\Op}{\mathcal{O}}
\newcommand{\Tr}{\mathop{\mathrm{Tr}}}

\newcommand{\order}[1]{\mathcal{O}(#1)}

\newcommand{\hateq}{\mathrel{\widehat=}}
\newcommand{\trans}{{\scriptscriptstyle\mathrm{T}}}

\newcommand{\cdott}{\mathord{\cdot}}

\ifx\genfrac\sdflkaj
\newcommand{\atopfrac}[2]{{{#1}\above0pt{#2}}}
\else
\newcommand{\atopfrac}[2]{\genfrac{}{}{0pt}{}{#1}{#2}}
\fi
\newcommand{\sfrac}[2]{{\textstyle\frac{#1}{#2}}}
\newcommand{\half}{\sfrac{1}{2}}
\newcommand{\ihalf}{\sfrac{i}{2}}

\newcommand{\indup}[1]{_{\mathrm{#1}}}

\newcommand{\rep}[1]{{\mathbf{#1}}}
\newcommand{\matr}[2]{\left(\begin{array}{#1}#2\end{array}\right)}
\newcommand{\alg}[1]{\mathfrak{#1}}
\newcommand{\grp}[1]{\mathrm{#1}}

\newcommand{\vect}[1]{\mathbf{#1}}
\newcommand{\mat}[1]{\mathbf{#1}}

\newcommand{\lrbrk}[1]{\left(#1\right)}
\newcommand{\bigbrk}[1]{\bigl(#1\bigr)}

\newcommand{\comm}[2]{[#1,#2]}

\newcommand{\state}[1]{\mathopen{|}#1\mathclose{\rangle}}

\newcommand{\nln}{\nonumber\\}
\newcommand{\nl}[1][0pt]{\nonumber\\[#1]&\hspace{-4\arraycolsep}&\mathord{}}

\newcommand{\earel}[1]{\mathrel{}&\hspace{-2\arraycolsep}#1\hspace{-2\arraycolsep}&\mathrel{}}
\newcommand{\eq}{\earel{=}}

\def\[{\begin{equation}}
\def\]{\end{equation}}
\def\<{\begin{eqnarray}}
\def\>{\end{eqnarray}}

\newcounter{enumlistcnt}
\renewcommand{\theenumlistcnt}{\protect\emph{\roman{enumlistcnt}}}

\makeatletter
\def\mr@ignsp#1 {\ifx\:#1\@empty\else #1\expandafter\mr@ignsp\fi}%
\newcommand{\multiref}[1]{\begingroup
\xdef\mr@no@sparg{\expandafter\mr@ignsp#1 \: }%
\def\mr@comma{}%
\@for\mr@refs:=\mr@no@sparg\do{\mr@comma\def\mr@comma{,}\ref{\mr@refs}}%
\endgroup}
\makeatother

\newcommand{\hypref}[2]{\ifx\href\asklfhas #2\else\href{#1}{#2}\fi}
\newcommand{\Secref}[1]{Section~\multiref{#1}}
\newcommand{\secref}[1]{Sec.~\multiref{#1}}

\newcommand{\Tabref}[1]{Table~\multiref{#1}}

\newcommand{\figref}[1]{Fig.~\multiref{#1}}
\renewcommand{\eqref}[1]{(\multiref{#1})}

\ifx\href\asklfhas\newcommand{\href}[2]{#2}\fi
\newcommand{\arxivno}[1]{\href{http://arxiv.org/abs/#1}{#1}}

\def\IR{\mathbb{R}}

\def\IZ{\mathbb{Z}}

\def\ialf{{\textstyle{\frac{i}{2}}}}

\def\id{\protect{{1 \kern-.28em {\rm l}}}}

\begin{document}


\thispagestyle{empty}
\begin{flushright}\footnotesize
\texttt{\arxivno{hep-th/0505187}}\\
\texttt{PUTP-2162}\\
\vspace{0.5cm}
\end{flushright}
\vspace{1.0cm}

\renewcommand{\thefootnote}{\fnsymbol{footnote}}
\setcounter{footnote}{0}

\begin{center}
{\Large\textbf{\mathversion{bold}
Beauty and the Twist:\\The Bethe Ansatz for Twisted ${\cal N}=4$ SYM
}\par} \vspace{1.5cm}

\textsc{N.~Beisert and R.~Roiban}\vspace{5mm}

\textit{Joseph Henry Laboratories\\
Princeton University\\
Princeton, NJ 08544, USA}\vspace{3mm}

\texttt{nbeisert,rroiban@princeton.edu}\\
\par\vspace{1cm}

\vfill

\textbf{Abstract}\vspace{5mm}

\begin{minipage}{12.7cm}
It was recently shown that the string theory duals of certain
deformations of the ${\cal N}=4$ gauge theory can be obtained by
a combination of T-duality transformations and coordinate shifts. 
Here we work out the corresponding procedure of twisting
the dual integrable spin chain and its Bethe ansatz.
We derive the Bethe equations 
for the complete twisted $\superN=4$ gauge theory 
at one and higher loops.
These have a natural generalization which we identify
as twists involving the Cartan generators 
of the conformal algebra. 
The underlying model appears to be a form of noncommutative deformation 
of $\superN=4$ SYM.
\end{minipage}

\vspace*{\fill}

\end{center}

\newpage
\setcounter{page}{1}
\renewcommand{\thefootnote}{\arabic{footnote}}
\setcounter{footnote}{0}


\section{Introduction}
\label{sec:Intro}

Over the years, the relation between gauge theories and string theory
has been an abundant source of understanding 
of both theories.
In the form of the AdS/CFT correspondence
\cite{Maldacena:1998re,Gubser:1998bc,Witten:1998qj} it
provides us with an explicit framework in which such information can
be extracted. One of the intriguing and fascinating developments is the
integrability of both the gauge theory dilatation operator 
\cite{Minahan:2002ve,Beisert:2003tq,Beisert:2003yb,Beisert:2004ry}
and of the world sheet sigma model \cite{Mandal:2002fs, Bena:2003wd, 
Vallilo:2003nx}.
Nevertheless, understanding the duality beyond
the near-BPS limit remains challenging, even within the set of
observables for which integrability plays a dominant role. 

The study of the deformations of ${\cal N}=4$ super-Yang-Mills theory (SYM) 
provides new controlled instances in which the duality may be 
tested. Given the reduced amount of symmetry, such a setup represents
a different way of departing from the near-BPS regime.
A natural place to start is provided by the exactly marginal
deformations of the theory \cite{Leigh:1995ep}. Some steps in this
direction have been taken in \cite{Roiban:2003dw,Berenstein:2004ys} 
where the Leigh-Strassler or $\beta$-deformation has been studied from the
standpoint of the one-loop dilatation operator of certain sectors 
of the theory.

More recently, in \cite{Lunin:2005jy} it was argued that a sequence of T-duality
transformations and coordinate shifts yields the 
supergravity duals to deformations of ${\cal N}=4$ SYM
which preserve all the Cartan generators of the superconformal
algebra. This allowed the explicit construction 
of the supergravity background dual to the $\beta$-deformed theory.   
It turns out that bosonic world sheet theory in 
this background exhibits a Lax pair \cite{Frolov:2005dj} similar to
the undeformed case \cite{Arutyunov:2004yx, Alday:2005gi}. 
Furthermore, the string Bethe equations derived from its
restriction to two world sheet fields agree with the thermodynamic
limit of the Bethe equations in the two-spin sector of the deformed
theory  \cite{Frolov:2005ty}. 

The algorithm of \cite{Lunin:2005jy} was used in \cite{Frolov:2005dj} 
for the construction of the supergravity background dual to a
three-parameter family of (generically) non-supersymmetric deformations of
${\cal N}=4$ SYM. The reasons implying the integrability of the
bosonic world sheet theory in the background dual to $\beta$-deformed 
SYM go through in this more general case as well. In fact, given that
the transformations leading to this background are well-defined in
string theory, we expect that the proof \cite{Berkovits:2004xu}
that the world sheet nonlocal charges survive quantization goes
through without major alterations.

It is then of great interest to analyze the field theory duals of
general deformations \cite{Lunin:2005jy, Frolov:2005dj}, which we will
generically refer to as ``twisted ${\cal N}=4$ SYM''. The gauge theory
duals to these backgrounds involve certain phase deformations of the 
terms appearing in the Lagrangian. It turns out that there are in fact
more possible deformations than those appearing in this string theory 
construction and it is interesting to identify those preserving the
integrability of the dilatation operator. From a different
perspective, the spin chain describing the dilatation operator of
${\cal N}=4$ SYM exhibits many integrable deformations similar in
spirit with those described by string theory, yet
different; it is interesting to learn which of them can be realized
within the class of twisted ${\cal N}=4$ theories.

In this paper we address these issues.
To construct the Bethe equations for the complete twisted ${\cal N}=4$
SYM we draw information from the sectors whose Hamiltonians can be
easily constructed from standard Feynman diagram calculations. In
\Secref{sec:ChTw} we start from the three-phase deformation introduced
in \cite{Frolov:2005dj} and show that the phases appearing while
reordering fields are determined by the phases appearing in the
reordering of fermions. Using this observation we construct an
operator whose eigenvalues are these phases, for any representation of
the reordered fields. Since this operator twists the spin chain
Hamiltonian describing the dilatation operator, it can be used to
twist the R-matrix of the chain as well. In \Secref{sec:DeformR} we
discuss this in detail and show that this operator generates 
an isomorphism
of the space of solutions of the Yang-Baxter equations. For the
sectors of ${\cal N}=4$ SYM described by spin chains in the fundamental
representation of a unitary group such twisted R-matrices reduce to
those discussed in \cite{Roiban:2003dw}. The largest sector with this
property is the $\alg{su}(2|3)$ sector of the undeformed theory,
see \cite{Beisert:2003ys} for a detailed account. 
In \Secref{sec:Nested} we review the derivation of 
the nested Bethe Ansatz for the $\alg{su}(2|3)$ sector of  
${\cal N}=4$ SYM and contrast it with the nested Bethe ansatz for the
twisted theory. 
The main conclusion of this analysis is that the Bethe
equations are the same as those obtained by twisting the 
diagonalized magnon S-matrix with the same twist operator used to twist the R-matrix.
We also discuss in principle the steps necessary to
apply the nested Bethe ansatz algorithm for other sectors of the
theory, not described by spin chains in the fundamental representation. 
Rather than following this line, in \Secref{sec:Complete} we make use
of the observation that the twist operator can be applied directly to the
magnon scattering matrix and discuss general flavor-dependent twists 
of the ${\cal N}=4$ spin chain. We show that the most general such
twist which has 
a Lagrangian realization within ${\cal N}=4$ SYM is the one
constructed in \cite{Frolov:2005dj}. Using different dual
presentations of the Dynkin diagram we identify various choices of the
twist parameters preserving supersymmetry and non-abelian global
symmetry. The same twist operation leads us to the conjecture that the 
higher-loop Bethe equations are a twisted form of those
conjectured in \cite{Beisert:2005fw} for the ${\cal N}=4$ SYM theory.
Interestingly, the consistency of this conjecture relies on the 
the same details as the compatibility of the twist with the Feynman 
rules at the one-loop level.
Last, we discuss deformations which
break Lorentz and conformal invariance. While it is not completely
clear what is the structure of the Lagrangian of the deformed field
theory, it is relatively straightforward to write the Bethe equations
for its dilatation operator;
we spell them out in \Secref{sec:NonCom}. Then, based on the naive
application of the construction \cite{Lunin:2005jy}, we discuss the
possible structure deformed Lagrangian as well as that of the
eigenvectors of the dilatation operator. \Secref{sec:Concl} contains 
further discussions.

\section{Charges and Twists}
\label{sec:ChTw}

As mentioned in the introduction, we are interested in
analyzing a certain class of integrable deformations of the 
${\cal N}=4$ spin chain 
and identifying those which correspond to rescaling terms of the 
Lagrangian by nontrivial constant phases.
We will draw information from examples involving 
closed subsectors of the theory, in which it is easy to explicitly
construct the dilatation operator starting from the Lagrangian. 
In this context we consider deformations which can be realized 
as a Moyal-like $*$-product based on the $\alg{su}(4)$ Cartan charges
of fields,  
introduced following \cite{Lunin:2005jy, Frolov:2005dj}. 
The gravity dual of the most general such deformation was constructed in 
\cite{Frolov:2005dj} and was also argued 
that the corresponding world sheet theory is
classically integrable.

The easiest sectors to analyze are those described by spin chains in
the fundamental representation of some unitary group, as the 
dilatation operator acts only by changing the
order of neighboring fields in gauge invariant operators, with certain
weights. The deformation of the Lagrangian implies that these weights
acquire nontrivial phases determined by the $\alg{su}(4)$-charges of the
fields being reordered. For each pair of fields they can be determined
from standard Feynman diagram calculations. However, due to the nature
of the deformation and the fact that the $\alg{su}(4)$-charges of all
fields are determined by those of the fermionic fields, the phase
obtained by reordering any two fields is the same as 
the phase obtained by reordering monomials constructed out of
fermions and carrying the same charges as the initial fields.
For example, the scalars $\Phi_A$  are in the $\rep{6}$ of
$\alg{su}(4)$ $\Phi_A\sim\Phi_{ab}$  and thus they have the same
$\alg{su}(4)$-charges as the fermion bilinears 
$\Psi_a\Psi_b$. 
We will use an $\alg{su}(3)$ ($\superN=1$) 
notation where a complex triplet of scalars $\phi_a$ is defined as
$\phi_a=\Phi_{a4}$ and $\bar\phi^a=\half\varepsilon^{abc}\Phi_{bc}$
and we will denote the triplet of fermions by $\psi_a=\Psi_a$, 
while $\chi=\Psi_4$ is the gluino.

This simple structure is somewhat modified in the sectors for which
the charges of the initial excitations are reorganized by the
interactions. Though more complicated, the deformation of the spin
chain Hamiltonian can still be determined from Feynman diagrams 
in terms of the
R-charges of the fundamental fermions. As we will see in detail 
later, the Bethe
ansatz presents us with a simple way of bypassing this slight 
complication. Indeed, in the undeformed theory, the
operators creating excitations corresponding to the simple roots of
the $\alg{psu}(2,2|4)$ in the ${\cal N}=4$ spin chain exhibit diagonal
scattering and thus no rearrangement of their $\alg{su}(4)$-charges. 

To summarize, we need to find the phases obtained by interchanging the
position of the fermions. They can be found easily by starting
from the deformed ${\cal N}=4$ Lagrangian, which is obtained by
replacing all commutators with deformed
commutators%
%
\[
\comm{X}{Y}_{\mat{C}}=e^{i \vect{q}_X\times \vect{q}_Y/2}\,XY-
e^{i\vect{q}_Y\times \vect{q}_X/2}\,YX
\]
or, alternatively, the structure constants are replaced with 
\[
f_{abc}X^bY^c~\mapsto~\bigbrk{\cos(\vect{q}_X\times \vect{q}_Y/2)\, f_{abc}
+i\sin(\vect{q}_X\times \vect{q}_Y/2)\, d_{abc}}X^bY^c~.
\]
Here we have introduced 
the antisymmetric $\mat{C}$-product
\[
\vect{q}_X\times \vect{q}_Y=
\vect{q}^\trans_X\mat{C}\vect{q}_Y=
C_{ab}q^a_X q^b_Y~.
\]
The $\alg{su}(4)$ Cartan charges $q^a_X$ (labeled by $a=1, 2, 3$) 
of the fundamental fields are given in \Tabref{tab:charges}.
\begin{table}\centering
$\begin{array}{|c|cccc|c|ccc|}
\hline
X&\psi_{1\alpha}&\psi_{2\alpha}&\psi_{3\alpha}&\chi_{\alpha}
&A_{\alpha\dot\alpha}&\phi_1&\phi_2&\phi_3\\
\hline
q^1_X&+\half&-\half&-\half&+\half&0&1&0&0\\
q^2_X&-\half&+\half&-\half&+\half&0&0&1&0\\
q^3_X&-\half&-\half&+\half&+\half&0&0&0&1\\
\hline
\end{array}$
\caption{Charges of the fundamental fields
under the Cartan generators of $\alg{su}(4)$.}
\label{tab:charges}
\end{table}
We define the matrix of phases $\mat{C}$ as 
\[
\label{Cmat}
\mat{C}=
\matr{ccc}{
\phantom{+}0 & -\gamma_3 & +\gamma_2\\
+\gamma_3 & \phantom{+}0 & -\gamma_1\\
-\gamma_2 & +\gamma_1 & \phantom{+}0}.
\]

When permuting two fermions $\Psi_a$ and $\Psi_b$ we pick up a phase
\[
\Psi_a\Psi_b\mapsto e^{i\vect{q}_a\times \vect{q}_b}\Psi_b\Psi_a
=
e^{iB_{ab}}\Psi_b\Psi_a
\]
where the phase matrix $\mat{B}$ 
in the basis $(\Psi_1,\Psi_2,\Psi_3,\Psi_4)$,
i.e.~$B_{ab}=\vect{q}_a\times \vect{q}_b$,
is given by
\[
\mat{B}=
\matr{cccc}{
0 & -\half(\gamma_1+\gamma_2) & +\half(\gamma_3+\gamma_1) & +\half(\gamma_2-\gamma_3)
\\
+\half(\gamma_1+\gamma_2) & 0 & -\half(\gamma_2+\gamma_3) & +\half(\gamma_3-\gamma_1)
\\
-\half(\gamma_3+\gamma_1) & +\half(\gamma_2+\gamma_3) & 0 & +\half(\gamma_1-\gamma_2)
\\
-\half(\gamma_2-\gamma_3)& -\half(\gamma_3-\gamma_1) & -\half(\gamma_1-\gamma_2)&  0
} .
\label{fundamentalphases}
\]
These phases can be used to reconstruct the phase matrix
for objects with multiple scalar indices like
the scalars $\Phi_{ab}\sim \Psi_a\Psi_b$ 
by summing up the contributions for permuting the
individual indices. For example, the phases among scalars 
$\phi_{a}=\Phi_{a4}\sim\Psi_a\Psi_4$ are given by
the original phase matrix
$\vect{q}_{a4}\times \vect{q}_{b4}=C_{ab}$.
Another example is the composite $\Psi_1\Psi_2\Psi_3\Psi_4$
for which all charges vanish.
Therefore the matrix $\mat{B}$ in \eqref{fundamentalphases}
annihilates the vector $(1,1,1,1)$
as can be easily verified.

\section{The Deformed R-matrix}
\label{sec:DeformR}

\begin{figure}[ht]\centering
$\displaystyle
\tilde{\mathcal{R}}_{ij}^{lk}
\equiv
\parbox{3.3cm}{\centering\includegraphics[width=3truecm]{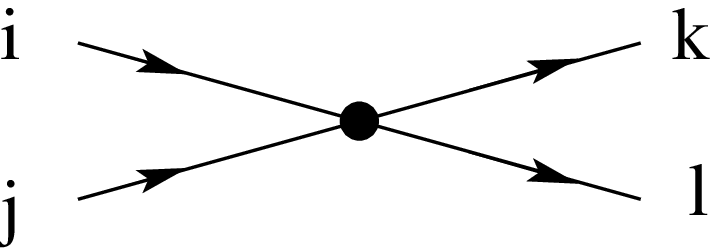}}
=
e^{i(\vect{q}_k\times \vect{q}_l - \vect{q}_i\times \vect{q}_j)/2}
\parbox{3.3cm}{\centering\includegraphics[width=3truecm]{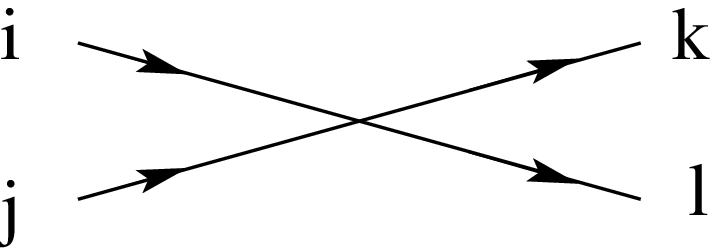}}
\equiv
e^{i(\vect{q}_k\times \vect{q}_l - \vect{q}_i\times \vect{q}_j)/2}
\mathcal{R}_{ij}^{lk}
$
\caption{Generic phase deformation of the R-matrix}
\label{pic_def}
\end{figure}

Let us consider a $\IZ_2$-graded set of states, labeled collectively by
$i$ and with the grade of the $i$-th state denoted by $[i]$. 
Let us also suppose that there exists a solution $\mathcal{R}$ of the
graded Yang-Baxter equation
\begin{equation}
\label{YBE}
(-)^{[j_2]([j_1]+[k_1])}
\mathcal{R}_{i_1i_2}^{j_1j_2}(u-v)
\mathcal{R}_{j_1i_3}^{k_1j_3}(u)
\mathcal{R}_{j_2j_3}^{k_2k_3}(v)=
(-)^{[j_2]([i_1]+[j_1])}
\mathcal{R}_{i_2i_3}^{j_2j_3}(v)
\mathcal{R}_{i_1j_3}^{j_1k_3}(u)
\mathcal{R}_{j_1j_2}^{k_1k_2}(u-v)
\end{equation}
which is labeled by these states. Assuming that the states carry 
conserved charges denoted by the charge vector $\vect{q}$, 
i.e.~$\vect{q}_i+\vect{q}_j=\vect{q}_k+\vect{q}_l$,
we will show\footnote{It turns out that this is a particular case of
Reshetikhin's construction of multiparametric quantum algebras
\cite{ReshetikhinTwist}.} that 
\[
\label{exp}
\tilde{\mathcal{R}}_{ij}^{lk}=
e^{i(\vect{q}_k\times \vect{q}_l - \vect{q}_i\times \vect{q}_j)/2}
\mathcal{R}_{ij}^{lk}
\]
is also a solution of the graded Yang-Baxter equation. 
Pictorially, this deformation can be represented as in \figref{pic_def}.

The number of such conserved charges depends, of course, on the
details of the undeformed R-matrix. Usually it may
be identified with the rank of the symmetry algebra 
(plus one for the conserved length of the spin chain).
In the simplest cases, for spin chains in the fundamental representation 
of a unitary group, there are as many such charges as spin orientations. 
Clearly however, this is not the generic situation and the number of
such charges is typically smaller than the dimension of the
representation labeling the R-matrix. 
Nevertheless, the deformation \eqref{exp} leads to a solution of the
Yang-Baxter equation.

The Yang-Baxter equation can be checked directly by noticing that,
after plugging  \eqref{exp} into \eqref{YBE},  the contribution of the
exponential factors is independent on the summation indices.  
\begin{figure}[ht]
\[\sum_{j_1,j_2,j_3}
\parbox{4.5cm}{\centering\includegraphics[height=4truecm]{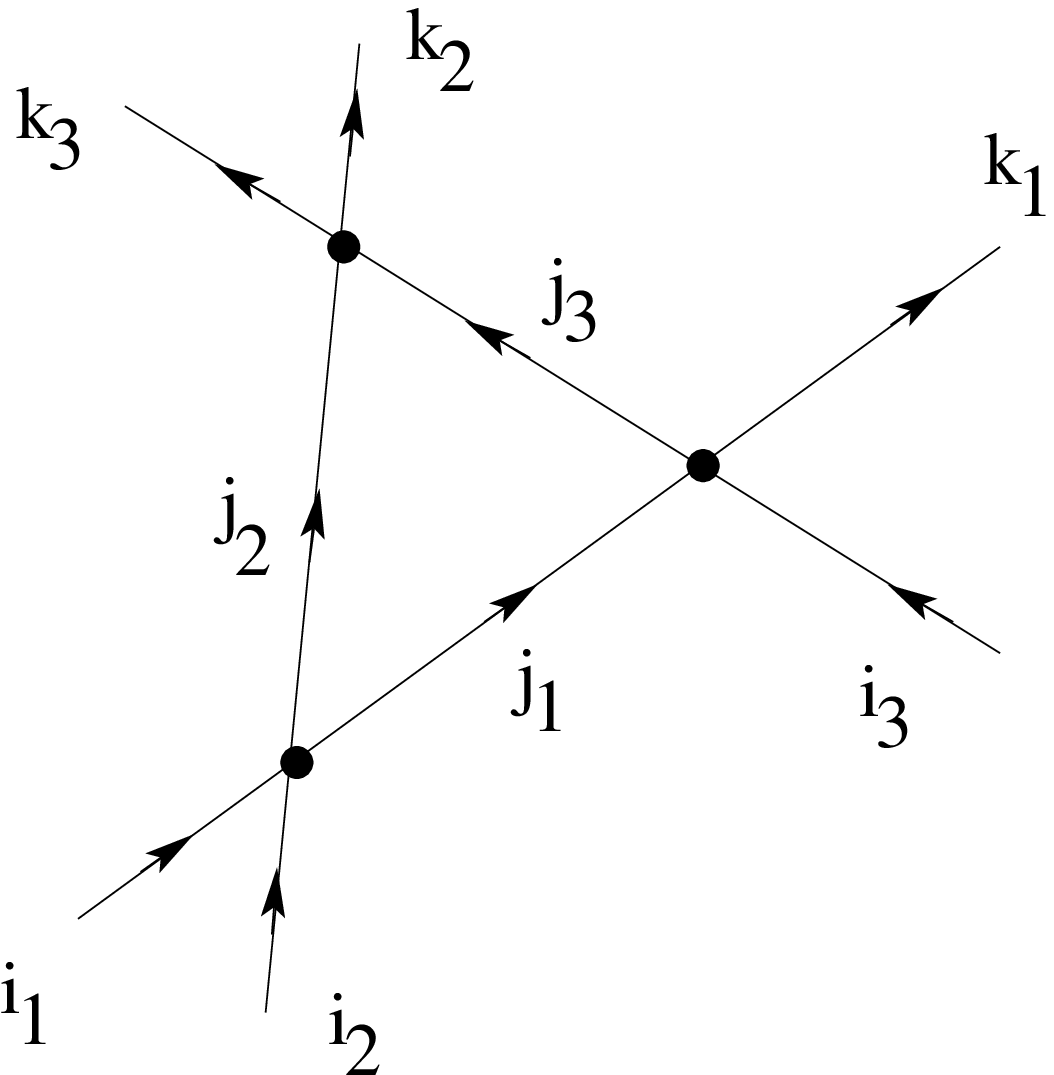}}
=
\sum_{j_1,j_2,j_3}
\parbox{4.5cm}{\centering\includegraphics[height=4truecm]{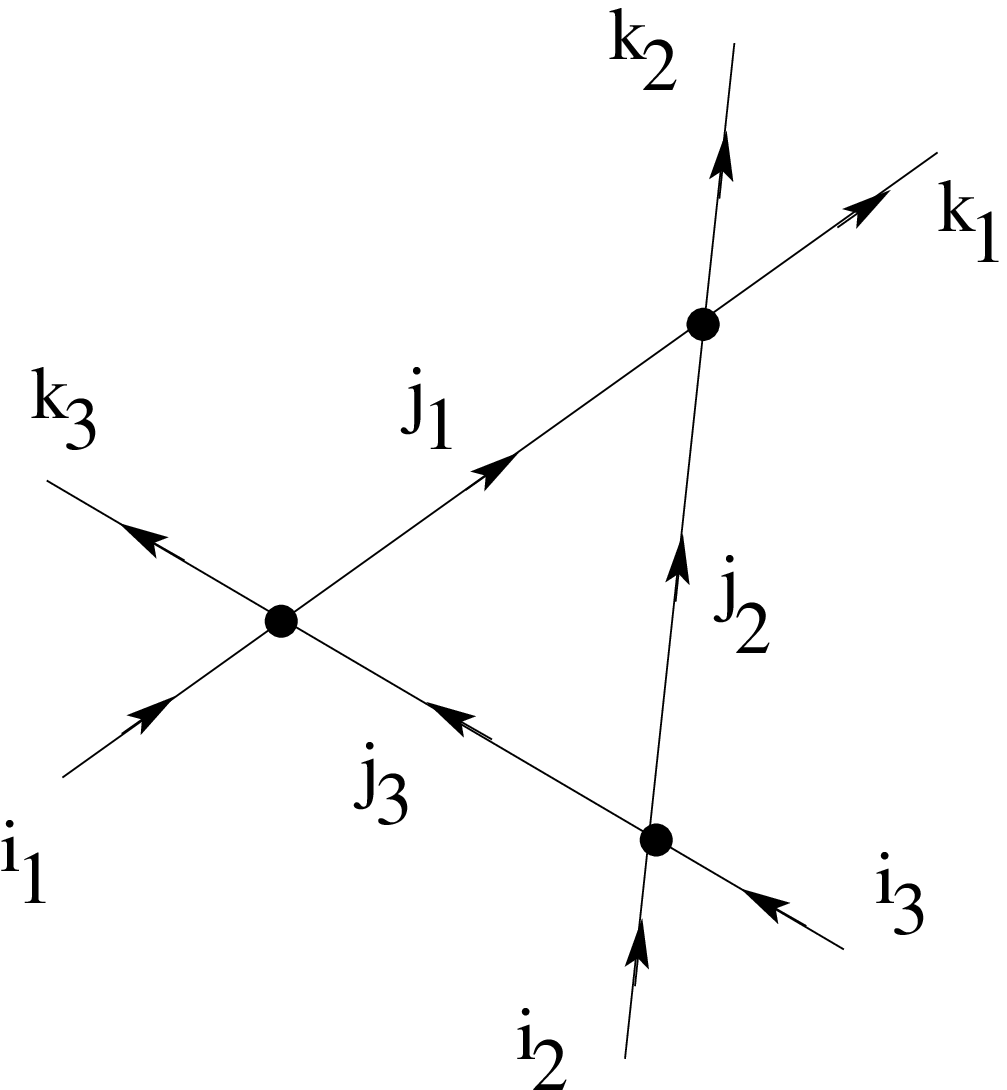}}
\]
\caption{The Yang-Baxter equation \label{YBEfig}}
\end{figure}
Explicitly, for fixed labels $j_1, j_2,j_3$ in \figref{YBEfig},
the phases on the two sides of the Yang-Baxter equation are:
\[\begin{array}[b]{c}
(\vect{q}_{j_2}\times \vect{q}_{j_1} - \vect{q}_{i_1}\times \vect{q}_{i_2})
+(\vect{q}_{j_3}\times \vect{q}_{k_1} - \vect{q}_{j_1}\times \vect{q}_{i_3})
+(\vect{q}_{k_3}\times \vect{q}_{k_2} - \vect{q}_{j_2}\times \vect{q}_{j_3}),
\\[0.3em]
(\vect{q}_{j_3}\times \vect{q}_{j_2} - \vect{q}_{i_2}\times \vect{q}_{i_3})
+(\vect{q}_{k_3}\times \vect{q}_{j_1} - \vect{q}_{i_1}\times \vect{q}_{j_3})
+(\vect{q}_{k_2}\times \vect{q}_{k_1} - \vect{q}_{j_1}\times \vect{q}_{j_2}),
\end{array}\]
both of which equal
\[
\vect{q}_{k_2}\times\vect{q}_{k_1}
+\vect{q}_{k_3}\times\vect{q}_{k_1}
+\vect{q}_{k_3}\times\vect{q}_{k_2}
-\vect{q}_{i_1}\times\vect{q}_{i_2}
-\vect{q}_{i_1}\times\vect{q}_{i_3}
-\vect{q}_{i_3}\times\vect{q}_{i_2}
.
\]
In showing this equality we only used the fact that the charges labeling
the states are conserved and that the $\times$-product is antisymmetric. 
We see that the combination of the deformations of the 
individual R-matrices is independent of the intermediate states and
equal on the two sides of the Yang-Baxter equation which is therefore
satisfied. 

More formally, one may notice that the deformation \eqref{exp} is
similar to the Moyal-deformation of field theories. {}From this
standpoint the Yang-Baxter equation may be interpreted as the equality
of two planar  one-loop Feynman diagrams.%
\footnote{In fact,
if the states labeling the R-matrix transform in the fundamental
representation of a unitary group, 
it is possible to write down a Lagrangian in which this is explicitly
realized. For example, one may consider a complex bosonic theory with
the four-point interactions given by 
$\mathcal{R}^{kl}_{ij}\phi_k\phi_l\bar\phi^i\bar\phi^j$.
}
Then, \cite{Filk:1996dm} implies that the
noncommutative deformation does not affect planar diagrams. Thus, it
follows that the deformation \eqref{exp} affects only the external
states in the Yang-Baxter equation.

The Hamiltonian of the deformed chain can be constructed following
standard rules. Starting from the deformed monodromy matrix 
\[
\mathcal{\tilde T}{}_{a;\alpha_1\dots\alpha_L}^{b;\beta_1\dots\beta_L}=
\tilde{\mathcal{R}}_{~~a~\alpha_L}^{b_{L-1}\beta_L}
\tilde{\mathcal{R}}_{b_{L-1}\alpha_{L-1}}^{b_{L-2}\beta_{L-1}}\dots
\tilde{\mathcal{R}}_{b_{2}\alpha_2}^{{b_1}\beta_2}
\tilde{\mathcal{R}}_{b_{1}\alpha_1}^{\,b\,\beta_1}\,
\exp\!\lrbrk{\scriptstyle{i\pi{\sum_{i=1}^L\sum_{j=1}^{i-1}
([\alpha_i]+[\beta_i])[\alpha_j]}}}\,,
\label{T10}
\]
the Hamiltonian is given by the logarithmic derivative of the transfer
matrix $\mathcal{\tilde T}(u)=(-)^{[a]} \mathcal{\tilde T}{}^a_a(u)$
\[
\mathcal{\tilde H}=-i\,\frac{d}{d u} \ln \mathcal{\tilde T}(u)\Big|_{u=u_*}
\]
where $u_*$ is the value of the rapidity at which
$\tilde{\mathcal{R}}$ becomes the graded permutation operator.
It is not difficult to see that using \eqref{exp} this leads to the 
following deformation of the Hamiltonian
\[\label{expham}
\mathcal{\tilde H}_{ij}^{kl}=
e^{i(\vect{q}_k\times \vect{q}_l - \vect{q}_i\times \vect{q}_j)/2}
\mathcal{H}_{ij}^{kl}.
\]

In the case in which the states labeling the undeformed Hamiltonian 
form the fundamental representation of some unitary group, it is easy
to see the effects of the deformation \eqref{exp}; since the undeformed
Hamiltonian is simply the sum between the identity operator and the
graded permutation operator, it follows that the deformed 
Hamiltonian is obtained by multiplying the graded permutation operator
by the same deformation as in \eqref{exp} while leaving the identity
operator unchanged.

\section{The Nested Bethe Ansatz in the $\alg{su}(2|3)$ Sector}
\label{sec:Nested}

Here we will derive the Bethe equations for a spin chain with 
$\alg{u}(2|3)$ symmetry and subsequently deform 
them according to the twist of the R-matrix from \Secref{sec:DeformR}.
This explicit analysis will suggest a generalization of
the twisted Bethe equations to arbitrary superalgebras.

The states of this spin chain correspond to
gauge theory operators 
composed from the fields
(with some unspecified permutation),
\[\label{states}
\state{n_1,\dots, n_5}
=(\phi_1)^{n_1}
(\phi_2)^{n_2}
(\phi_3)^{n_3}
(\chi_{1})^{n_4}
(\chi_{2})^{n_5}
+\ldots\,,
\]
see \cite{Beisert:2003ys} 
for a detailed account of this spin chain in connection with gauge theory.
The field $\phi_1$ will be considered the vacuum and
all the other fields are excitations.

\subsection{Undeformed case}

To track down the effects of the deformation \eqref{exp} on the Bethe
equations let us first review the diagonalization of the transfer
matrix of the $\alg{su}(2|3)$ sector of ${\cal N}=4$ SYM which proceeds in 
the standard way, following the algorithm of the nested Bethe Ansatz. 

The spin chain of the $\alg{su}(2|3)$ sector of ${\cal N}=4$ SYM is
described by the R-matrix in the fundamental representation 
of $\alg{u}(2|3)$:%
\footnote{At one loop we may take the $\alg{u}(1)$ factor 
of $\alg{u}(2|3)$ to represent the length of the spin chain.}
\[
\mathcal{R}(u)=\frac{1}{u+i}\left(u\,\mathcal{I} + i\mathcal{P}\right)
\label{fR}
\]
where $\mathcal{I}$ is the identity and $\mathcal{P}$ is the graded permutation operator 
\[
\mathcal{I}{}_{ij}^{kl}=\delta_i^k\delta_j^l,
\,\qquad
\mathcal{P}{}_{ij}^{kl}=(-)^{[k][l]}\delta_i^l\delta_j^k
\]
and can be expressed in terms of the generators of $\alg{u}(2|3)$ and the
Cartan-Killing metric:
\[
\mathcal{P}=\sum_{i, j} \, (-)^{[i]+[j]}\,e_i^j\otimes e_j^i
\]

Picking a vacuum $|0\rangle$ 
corresponding to $(\phi^1){}^L$,
the relevant commutation relations following from the (graded) Yang-Baxter
equation are:
\begin{eqnarray}
\mathcal{A}(u)\,
\mathcal{B}^{i_1}(v)\eq 
f(v-u)\,
\mathcal{B}^{i_1}(v)\,
\mathcal{A}(u)-
g(v-u)\,
\mathcal{B}^{i_1}(u)\,
\mathcal{A}(v),
\label{A0}\\
\mathcal{D}_{k_1}^{i_1}(u)\,
\mathcal{B}^{i_2}(v)
\eq
{(-)^{[k_1][j_2]}\,
r_{j_2j_1}^{i_1i_2}(u-v)}\,
f(u-v)\,
\mathcal{B}^{j_2}(v)\,
\mathcal{D}_{k_1}^{j_1}(u)
\nl
-(-)^{[k_1][i_1]}\,
g(u-v)\,
\mathcal{B}^{i_1}(v)\,
\mathcal{D}_{k_1}^{i_2}(u),
\label{D0}
\\
\label{BB0}
\mathcal{B}^{i_1}(u)\,
\mathcal{B}^{i_2}(v)\eq 
r^{i_1i_2}_{j_2j_1}(u-v)\,
\mathcal{B}^{j_2}(v)\,
\mathcal{B}^{j_1}(u),
\end{eqnarray}
where
\begin{equation}
r^{i_1i_2}_{j_1j_2}=(-)^{[j_1][j_2]}\mathcal{R}^{i_1i_2}_{j_2j_1}
~,~~~~~
g(u)=\frac{i}{u}~,
~~~~~
f(u)=\frac{u+i}{u}
\end{equation}
and the remaining labels $i, j=1,\ldots,4$ enumerate the fields
$\phi_2,\phi_3,\chi_1,\chi_2$.
Making the ansatz that a state 
with $n=n_2+n_3+n_4+n_5$ excitations is given by
\begin{equation}
|n_1,\dots, n_5\rangle=
f_{i_1\dots i_n}\mathcal{B}^{i_1}(u_{1, 1})\dots
\mathcal{B}^{i_n}(u_{1,n}) 
\state{0}
\label{state}
\end{equation}
and using \eqref{A0}-\eqref{BB0}, the diagonalization of transfer matrix
constructed out of $\mathcal{R}$ is reduced to the diagonalization of the
transfer matrix constructed out of $r$ which also satisfies the
Yang-Baxter equation and the (wave) functions $f_{i_1\dots i_n}$ are
the eigenvectors of this (reduced) transfer matrix.

Repeating these steps four times leads to the Bethe equations:
\begin{eqnarray}
1\eq~~\,\left(\frac{u_{1,k}-\ialf}{u_{1,k}+\ialf}\right)^L~~\,
\prod_{\textstyle\atopfrac{j=1}{j\neq k}}^{K_1}
\frac{u_{1,k}-u_{1,j}+i}{u_{1,k}-u_{1,j}-i}\,
\prod_{j=1}^{K_2}
\frac{u_{1,k}-u_{2,j}-\ialf}{u_{1,k}-u_{2,j}+\ialf}\,,
\cr
1\eq
\prod_{j=1}^{K_1}
\frac{u_{2,k}-u_{1,j}-\ialf}{u_{2,k}-u_{1,j}+\ialf}
\prod_{\textstyle\atopfrac{j=1}{j\neq k}}^{K_2}
\frac{u_{2,k}-u_{2,j}+i}{u_{2,k}-u_{2,j}-i}
\prod_{j=1}^{K_3}
\frac{u_{2,k}-u_{3,j}-\ialf}{u_{2,k}-u_{3,j}+\ialf}\,,
\cr
1\eq
\prod_{j=1}^{K_2}
\frac{u_{3,k}-u_{2,j}-\ialf}{u_{3,k}-u_{2,j}+\ialf}
\mathop{\phantom{\prod_{\textstyle\atopfrac{j=1}{j\neq k}}^{K_3}
\frac{u_{3,k}-u_{3,j}+i}{u_{3,k}-u_{3,j}-i}}}
\prod_{j=1}^{K_4}
\frac{u_{3,k}-u_{4,j}+\ialf}{u_{3,k}-u_{4,j}-\ialf}\,,
\cr
1\eq
\prod_{j=1}^{K_3}
\frac{u_{4,k}-u_{3,j}+\ialf}{u_{4,k}-u_{3,j}-\ialf}
\prod_{\textstyle\atopfrac{j=1}{j\neq k}}^{K_4}
\frac{u_{4,k}-u_{4,j}-i}{u_{4,k}-u_{4,j}+i}
\end{eqnarray}
Furthermore, the cyclicity constraint
\[
1=\prod_{k=1}^{K_1}\frac{u_{1,k}+\ialf}{u_{1,k}-\ialf}
\]
ensures that the operators described by the states \eqref{states}
are compatible with taking a color trace.
Here we have introduced the numbers of Bethe roots $K_j$
specified by
\[\label{eq:ExRel}
K_j=n_{j+1}+\ldots+n_5,\qquad
n_j=K_{j-1}-K_{j}.
\]
Finally, the anomalous dimension of a state reads
\[\label{eq:ano}
\delta
D=g^2\sum_{k=1}^{K_1}\lrbrk{\frac{i}{u_{1,k}+\ialf}
-\frac{i}{u_{1,k}-\ialf}}
+\order{g^4}~.
\]

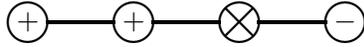
\begin{figure}\centering
\begin{minipage}{140pt}
\setlength{\unitlength}{1pt}%
\small\thicklines%
\begin{picture}(140,20)(-10,-10)
\put(  0,00){\circle{15}}%
\put(  7,00){\line(1,0){26}}%
\put( 40,00){\circle{15}}%
\put( 47,00){\line(1,0){26}}%
\put( 80,00){\circle{15}}%
\put( 87,00){\line(1,0){26}}%
\put(120,00){\circle{15}}%
\put( 75,-5){\line(1, 1){10}}%
\put( 75, 5){\line(1,-1){10}}%
\put( 00,00){\makebox(0,0){$+$}}%
\put( 40,00){\makebox(0,0){$+$}}%
\put(120,00){\makebox(0,0){$-$}}%
\end{picture}
\end{minipage}

\caption{Dynkin diagram of $\alg{su}(2|3)$.
The signs in the white nodes indicate
the sign of the diagonal elements of the
Cartan matrix.}
\label{fig:Dynkin23}
\end{figure}

We can write the Bethe equation and the momentum constraint in a concise form
\[\label{eq:Bethe}
\lrbrk{
\frac{u_{j,k}-\ihalf V_j}{u_{j,k}+\ihalf V_j}}^L
\mathop{\prod_{j'=1}^N\prod_{k'=1}^{K_{j'}}}_{(j',k')\neq(j,k)}
\frac{u_{j,k}-u_{j',k'}+\ihalf M_{j,j'}}{u_{j,k}-u_{j',k'}-\ihalf M_{j,j'}}
=1~,\qquad
\prod_{j=1}^N
\prod_{k=1}^{K_j}
\frac{u_{j,k}+\ihalf V_j}{u_{j,k}-\ihalf V_j}
=1~.
\]
Here $N=4$ is the rank of the symmetry algebra, 
$V_j=(1,0,0,0)$ are the Dynkin labels of the spin representation
and $\mat{M}$ is the symmetric Cartan matrix, cf.~\figref{fig:Dynkin23}.
\[
\mat{M}=\matr{cccc}{+2&-1&&\\-1&+2&-1&\\&-1&&+1\\&&+1&-2}~.
\]
This is the universal form of Bethe equations for standard quantum
spin chains with arbitrary symmetry algebra due to
\cite{Reshetikhin:1983vw, Reshetikhin:1985vd, Ogievetsky:1986hu}.

\subsection{Deformed case}
\label{sec:Deform23}

It is not hard to apply the nested Bethe ansatz 
to the $\alg{u}(2|3)$ R-matrix deformed as in \eqref{exp}. 
Since the graded permutation operator preserves the
charge vectors of the incoming and outgoing excitations while only
exchanging them,  the deformation acts trivially it.
We see therefore that only the term in \eqref{fR} involving 
the identity operator is deformed; consequently, the R-matrix
describing the deformed $\alg{u}(2|3)$ spin chain is 
\[
\tilde{\mathcal{R}}(u)_{12}
=\frac{1}{u+i}\bigbrk{u\,e^{-i \vect{q}_1\times
\vect{q}_2}\,\mathcal{I}_{12} + i{\cal P}_{12}}
~~~~~~~~ 
\tilde{\mathcal{R}}(u){}_{ij}^{kl}
=\frac{1}{u+i}\bigbrk{u\,e^{-iB_{ij}}\,
\mathcal{I}{}_{ij}^{kl} + i{\cal P}{}_{ij}^{kl}
}
\label{fRd}
\]
It is worth emphasizing that, from the standpoint of the spin chain
for the undeformed $\alg{u}(2|3)$ algebra, any choice for the matrix
$\mat{B}$ is allowed. We will take this standpoint in this section and
will not commit to a particular form for its matrix elements. 
In section \ref{recapU23} we specialize to the $\mat{B}$-matrix following
from the discussion in section \ref{sec:ChTw}.
In some sense this corresponds to 
removing the central $\alg{u}(1)$ in $\alg{u}(2|3)$
(which is not actually a symmetry of $\superN=4$ SYM)
from the set of Cartan generators used for twisting the  
spin chain.

The simplicity of the deformation \eqref{fRd} implies that the 
modification of the commutation relations following from the
Yang-Baxter equations are also minimal: 
\begin{eqnarray}
\mathcal{A}(u)\,\mathcal{B}^{i_1}(v)\eq 
e^{-B_{i_1 1}}
\left[f(v -u)\,\mathcal{B}^{i_1}(v )\,\mathcal{A}(u)-
g(v -u)\,\mathcal{B}^{i_1}(u)\,\mathcal{A}(v )\right]~,
\label{A0d}\\
\mathcal{D}_{k_1}^{i_1}(u)\,\mathcal{B}^{i_2}(v)\eq 
e^{-B_{k_1 1}}\bigl[
{(-)^{[k_1][j_2]}\,r_{j_2j_1}^{i_1i_2}(u-v )}\,f(u-v )\,
\mathcal{B}^{j_2}(v )\,\mathcal{D}_{k_1}^{j_1}(u)
\nl
\qquad\qquad\qquad\qquad\qquad
-(-)^{[k_1][i_1]}\,g(u-v )\,
\mathcal{B}^{i_1}(v )\,\mathcal{D}_{k_1}^{i_2}(u)
\bigr]~,
\label{D0d}
\\
\label{BB0d}
\mathcal{B}^{i_1}(u)\,\mathcal{B}^{i_2}(v )\eq 
r^{i_1i_2}_{j_2j_1}(u-v )\,
\mathcal{B}^{j_2}(v )\,\mathcal{B}^{j_1}(u)~.
\end{eqnarray}
This in turn leads to inserting phases in the Bethe equations
following \eqref{fundamentalphaseseg}. 
\begin{eqnarray}
1\eq
e^{i (n_1+n_2) B_{21}}
e^{in_3 (B_{23}+B_{31})}
e^{in_4 (B_{24}+B_{41})}
e^{in_5 (B_{25}+B_{51})}
\nl\times
~~\,\left(\frac{u_{1,k}-\ialf}{u_{1,k}+\ialf}\right)^L~~\,
\prod_{\textstyle\atopfrac{j=1}{j\neq k}}^{K_1}
\frac{u_{1,k}-u_{1,j}+i}{u_{1,k}-u_{1,j}-i}\,
\prod_{j=1}^{K_2}
\frac{u_{1,k}-u_{2,j}-\ialf}{u_{1,k}-u_{2,j}+\ialf}\,,
\cr
1\eq
e^{in_1(B_{31}+B_{12})}
e^{i(n_2+n_3) B_{32}}
e^{in_4 (B_{34}+B_{42})}
e^{in_5 (B_{35}+B_{52})}
\nl\times
\prod_{j=1}^{K_1}
\frac{u_{2,k}-u_{1,j}-\ialf}{u_{2,k}-u_{1,j}+\ialf}
\prod_{\textstyle\atopfrac{j=1}{j\neq k}}^{K_2}
\frac{u_{2,k}-u_{2,j}+i}{u_{2,k}-u_{2,j}-i}
\prod_{j=1}^{K_3}
\frac{u_{2,k}-u_{3,j}-\ialf}{u_{2,k}-u_{3,j}+\ialf}\,,
\cr
1\eq
e^{in_1(B_{41}+B_{13})}
e^{in_2 (B_{42}+B_{23})}
e^{i(n_3+n_4) B_{43}}
e^{in_5 (B_{45}+B_{53})}
\nl\times
\prod_{j=1}^{K_2}
\frac{u_{3,k}-u_{2,j}-\ialf}{u_{3,k}-u_{2,j}+\ialf}
\mathop{\phantom{\prod_{\textstyle\atopfrac{j=1}{j\neq k}}^{K_3}
\frac{u_{3,k}-u_{3,j}+i}{u_{3,k}-u_{3,j}-i}}}
\prod_{j=1}^{K_4}
\frac{u_{3,k}-u_{4,j}+\ialf}{u_{3,k}-u_{4,j}-\ialf}\,,
\cr
1\eq
e^{in_1(B_{51}+B_{14})}
e^{in_2 (B_{52}+B_{24})}
e^{in_3 (B_{53}+B_{34})}
e^{i(n_4+n_5) B_{54}}
\nl\times
\prod_{j=1}^{K_3}
\frac{u_{4,k}-u_{3,j}+\ialf}{u_{4,k}-u_{3,j}-\ialf}
\prod_{\textstyle\atopfrac{j=1}{j\neq k}}^{K_4}
\frac{u_{4,k}-u_{4,j}-i}{u_{4,k}-u_{4,j}+i}\,.
\label{eq:BetheTwist23}
\end{eqnarray}
Let us now consider the effect of the deformation on the local charge 
eigenvalues $Q_r$ which appear in the expansion of the 
transfer matrix eigenvalue $T(u)$ around $u=u_\ast$
\[
T(u)=\exp i\sum_{r=1}^\infty (u-u_\ast)^{r-1} Q_r
\]
Now recall the fact that the eigenvalues
of the transfer matrix are given by a sum of five terms,
one for each component of the fundamental 
representation of $\alg{u}(2|3)$.
These stem from commuting the ${\mathcal A}$ and diagonal entries of
the ${\mathcal D}$ operators 
past the creation operators in the state \eqref{state}. 
Furthermore, these five terms are proportional to the eigenvalues 
of the ${\mathcal A}$ and ${\mathcal D}$ operators corresponding 
to the vacuum state $\state{0}$, the latter vanishing for $u=u_*$. 
Thus, the nontrivial contribution to the eigenvalues 
of the Hamiltonian come from a single term. 
The deformation is reflected in this term through a
multiplicative $u$-independent phase factor.
The expression for the total momentum $Q_1$ picks up this factor and
yields the deformed cyclicity constraint
\[
1=
e^{in_2B_{21}}
e^{in_3B_{31}}
e^{in_4B_{41}}
e^{in_5B_{51}}
\prod_{j=1}^{K_1}\frac{u_{1,j}+\ialf}{u_{1,j}-\ialf}\,.
\label{eq:MomTwist23}
\]
Beyond that, it has no effect on any of 
the expressions of the local conserved charges,
in particular on the anomalous dimension $\delta D=g^2 Q_2$.
Dependence on the deformation parameters comes 
only through the solutions of the deformed Bethe equations.

We can now translate all the numbers of individual fields 
$n_a$ to the excitation numbers $K_j$ using \eqref{eq:ExRel}.
It is natural and convenient to set $K_0=L$
with $L$ the length of the spin chain.
We therefore introduce the vector
\[
\vect{K}=(L\mathpunct{|}K_1,\ldots, K_N)
\]
where $N=4$ is the rank of the symmetry algebra.
The elements of the matrix $\mat{B}$ then group 
naturally in a new $(1+N)\times(1+N)$ 
antisymmetric matrix $\mat{A}$ defined by 
\[\label{eq:TranslatePhases}
A_{j,j'}=B_{j,j'}-B_{j,j'+1}-B_{j+1,j'}+B_{j+1,j'+1},
\]
Note that $j,j'=0,\ldots, N$ 
and we assume $B_{0,\ast}=B_{\ast,0}=0$.

The above twisted Bethe equations \eqref{eq:BetheTwist23}
can now be written conveniently as
\[\label{eq:TwistBethe}
e^{i(\mat{A}\vect{K})_j}
\lrbrk{
\frac{u_{j,k}-\ihalf V_j}{u_{j,k}+\ihalf V_j}}^L
\mathop{\prod_{j'=1}^N\prod_{k'=1}^{K_{j'}}}_{(j',k')\neq(j,k)}
\frac{u_{j,k}-u_{j',k'}+\ihalf M_{j,j'}}{u_{j,k}-u_{j',k'}-\ihalf
M_{j,j'}} 
=1
\]
and the twisted zero-momentum constraint 
\eqref{eq:MomTwist23} reads
\[\label{eq:TwistMomentum}
e^{i(\mat{A}\vect{K})_0}
\prod_{j=1}^N
\prod_{k=1}^{K_j}
\frac{u_{j,k}+\ihalf V_j}{u_{j,k}-\ihalf V_j}
=1.
\]
Note that we have used a compact vector notation to write 
the total phases as
\[(\mat{A}\vect{K})_j=A_{j,0}L+\sum_{j'=1}^N A_{j,j'}K_{j'},\qquad
(\mat{A}\vect{K})_0=\sum_{j=1}^N A_{0,j}K_j.
\]
%

\subsection{Deformations of chains in arbitrary representations}
\label{sec:mostgeneral}

We have discussed in detail a class of deformations of spin chains in 
the fundamental representation of some unitary group focusing on the
largest subsector of ${\cal N}=4$ SYM with this property -- the 
$\alg{su}(2|3)$ sector --  and have seen how the deformation affects
the structure of the Bethe equations. However, in the study of the
dilatation operator of gauge theories more general spin chains appear,
most notably the spin chain based on the self-conjugate doubleton
representation of $\alg{psu}(2,2|4)$,  describing the dilatation
operator of the full ${\cal N}=4$ SYM theory
\cite{Beisert:2003jj,Beisert:2003yb}.  
It is natural to ask whether our discussion can be extended to such
more general chains and if so how many of the deformation parameters 
survive.

Following the arguments in \secref{sec:DeformR} it is not hard to
see that all deformation parameters appearing at the level of the
R-matrix in the fundamental representation extend without
restrictions to more general representations. Indeed, the form of the
Yang-Baxter equation is independent of the dimension 
of the space of states
on the sites of the chain. Furthermore, we have shown on general
grounds that \eqref{exp} yields solutions of this equation. 

Starting from the fundamental representation of $\alg{u}(2,2|4)$, we
see that there are at most $28$ possible deformation parameters; they
form the $8\times 8$ analog of the antisymmetric matrix $\mat{C}$ in
equation \eqref{Cmat} which in turn determines the (possibly
infinite-dimensional) $\mat{B}$-matrix deforming the R-matrix and
thus the spin chain Hamiltonian. It is worth emphasizing that some 
of these deformations break Lorentz and conformal symmetry. We will
return to this in \secref{sec:NonCom}.

With this input, the diagonalization proceeds following the standard
algorithm. First one solves the Yang-Baxter equation for 
the ${\tilde R}$-matrix acting on the
tensor product of the fundamental representation and the
representation of interest. {}From \secref{sec:DeformR} it
follows that this is again a twisted form of the analogous R-matrix
in the undeformed theory. Since the sizes of the two Hilbert spaces
are now different, the corresponding $\mat{B}$-matrix
\eqref{fundamentalphases} is now a rectangular matrix, still
determined however by the same 28-parameter $\mat{C}$-matrix.
Then, the monodromy matrix with the auxiliary Hilbert 
space in the fundamental representation can be immediately constructed
and diagonalized. Due to the Yang-Baxter equation its eigenvectors are  
also the eigenvectors of the monodromy matrix with the auxiliary
Hilbert space in the same representation as the physical sites. 
Thus, the resulting Bethe equations determine the eigenvalues of the 
dilatation operator in the representation of interest.

It is certainly possible to follow this procedure and diagonalize the 
full spin chain for the deformed ${\cal N}=4$ SYM theory. Various
complications arise from the requirement that resulting eigenstates 
belong to representations of $\alg{psu}(2,2|4)$; for example, the
vacuum which may appear natural from the standpoint of the fundamental
representation of the $\alg{u}(2,2|4)$ does not satisfy this requirement. 
In the following sections  we will obtain the deformed Bethe equations
by simpler methods. Using those results one may go back and, using
the discussion above together with techniques of \cite{Dolan:2004ys}, 
one may reconstruct the one-loop dilatation operator of the full deformed
${\cal N}=4$ SYM theory.

\section{Flavor Deformations for $\superN=4$ SYM}
\label{sec:Complete}

In this section we apply the above findings 
to the complete spin chain model
involving all scalars, fermions, field strengths as well as
their covariant derivatives.

\subsection{The $\alg{su}(2|3)$ sector}
 \label{recapU23}

Let us start with the $\alg{su}(2|3)$ sector
whose field content is 
$(\phi_1,\phi_2,\phi_3\mathpunct{|}\chi_{1},\chi_{2})$.
Using the results of \secref{sec:Deform23} it is now straightforward
to obtain twist matrices.
In terms of the Cartan charges this set of fields is equivalent to the
vector  
$(\Psi_1\Psi_4,\Psi_2\Psi_4,\Psi_3\Psi_4\mathpunct{|}\Psi_{4},\Psi_{4})$.
The matrix of phases $B_{j,j'}=\vect{q}_j\times \vect{q}_{j'}$ 
can now be assembled from \eqref{fundamentalphases}
\[
\mat{B}=
\matr{ccc|cc}{
 \phantom{+}    0    & -\gamma_3  & +\gamma_2 & 
            \half(\gamma_2-\gamma_3) & \half(\gamma_2-\gamma_3)\\
+\gamma_3 &     \phantom{+}  0   &  -\gamma_1 & 
            \half(\gamma_3-\gamma_1) & \half(\gamma_3-\gamma_1)\\
-\gamma_2 & +\gamma_1 & \phantom{+}    0     & 
             \half(\gamma_1-\gamma_2) & \half(\gamma_1-\gamma_2)
\\\hline
\half(\gamma_3-\gamma_2)& 
\half(\gamma_1-\gamma_3)&
\half(\gamma_2-\gamma_1)& 0& 0\\
\half(\gamma_3-\gamma_2)&
\half(\gamma_1-\gamma_3)&
\half(\gamma_2-\gamma_1)& 0& 0}.
\label{fundamentalphaseseg}
\]
Consequently, the phase matrix for the Bethe equations is
obtained using \eqref{eq:TranslatePhases}
\[
\mat{A}=
\matr{c|cccc}{
 \phantom{+}    0    & -\gamma_3  & +\gamma_2+\gamma_3 & -\half\gamma_2-\half\gamma_3 & 0
\\\hline
+\gamma_3 &     \phantom{+}  0   &  -\gamma_1-\gamma_2-\gamma_3 & +\half\gamma_1+\half\gamma_2+\gamma_3  & 0
\\
-\gamma_2-\gamma_3 & +\gamma_1+\gamma_2+\gamma_3 & \phantom{+}    0     &  -\half\gamma_2-\half\gamma_3 & 0
\\
+\half\gamma_2+\half\gamma_3& 
-\half\gamma_1-\half\gamma_2-\gamma_3&
+\half\gamma_2+\half\gamma_3& 0& 0\\
0&0&0&0&0}.
\label{eq:Twistu23}
\]

The phases in the upper-left $3\times 3$ block 
agree with the phases obtained in \cite{Berenstein:2004ys}
in the supersymmetric case $\gamma_1=\gamma_2=\gamma_3$.
Also the Bethe equation for the $\alg{su}(2)$ sector 
(upper-left $2\times 2$ block) agrees with \cite{Frolov:2005ty}.

\subsection{The $\alg{so}(6)$ sector}

Even though it appears that this sector suffers from the complication
we mentioned in section \ref{sec:ChTw} -- that the $\alg{su}(4)$
charges are reorganized by the Hamiltonian -- it is in fact easy to
see that the apparently problematic term in the R-matrix -- the trace
operator --  is undeformed by \eqref{exp}. Consequently, it should not
be too complicated to apply the nested Bethe ansatz and find the
deformed Bethe equations. Here however we will not follow this route,
but rather start from the undeformed Bethe equations and interpret
them in terms of the magnon scattering matrix, which is in turn
deformed following \eqref{exp}.

For reasons which will become clear later we will assume 
the phase matrix determining the commutation of the 
the three complex scalars $(\phi_1,\phi_2,\phi_3)$
to be
\[
\mat{B}=\matr{ccc}{
0&+\delta_1-\delta_3&-\delta_1-\delta_3\\
-\delta_1+\delta_3&0&+\delta_1+2\delta_2+\delta_3\\
+\delta_1+\delta_3&-\delta_1-2\delta_2-\delta_3&0}.
\]
This is equivalent to the upper left block in
\eqref{fundamentalphaseseg}, but with a different 
parametrization $\delta_j$ of the phases $\gamma_j$.
The new phases are chosen such that the energies 
are invariant under a shift of any $\delta_j$ by $2\pi$.
For commuting two conjugate scalars $(\bar\phi^1,\bar\phi^2,\bar\phi^3)$
the same matrix $\mat{B}$ applies, while a mixed commutator 
is determined by $-\mat{B}$.
This determines all the phases for the $\alg{so}(6)$ spin chain.

Now we have to convert the twist matrix of the fields
into a twist matrix for the excitations of the Bethe ansatz. 
For the rank-three algebra $\alg{so}(6)$, there are
three types of excitations, 
let us denote their creation operators by $\mathcal{B}_1,\mathcal{B}_2,\mathcal{B}_3$. 
As the vacuum we will choose the field $\phi_1$.
The actions of the three excitations are as follows
\[\begin{array}{ccc}
\mathcal{B}_1:& \phi_2\mapsto\bar\phi^3, & \phi_3\mapsto\bar\phi^2,
\\[0.3em]
\mathcal{B}_2:& \phi_1\mapsto\phi_2, & \bar\phi^2\mapsto\bar\phi^1,
\\[0.3em]
\mathcal{B}_3:& \phi_2\mapsto\phi_3, & \bar\phi^3\mapsto\bar\phi^2.
\end{array}
\]
We would now like to transform the twist matrix to the basis
$(\phi_1\mathpunct{|}\mathcal{B}_1,\mathcal{B}_2,\mathcal{B}_3)$ which is, 
in terms of the charges, equivalent to 
$(\phi_1\mathpunct{|}\bar\phi^2\bar\phi^3,\bar\phi^1\phi_2,\bar\phi^2\phi_3)$.
We consequently add or subtract the rows and columns according to the
composite nature of the excitations $\mathcal{B}_k$ in terms of the $\phi_a$.
The twist matrix for magnons in the Bethe equations is
\[\label{eq:TwistSO6}
\mat{A}=\matr{c|ccc}{
0&+2\delta_3&+\delta_1-\delta_3&-2\delta_1\\\hline
-2\delta_3&0&+\delta_1+2\delta_2+3\delta_3&-2\delta_1-4\delta_2-2\delta_3\\ 
-\delta_1+\delta_3&-\delta_1-2\delta_2-3\delta_3&0&+3\delta_1+2\delta_2+\delta_3\\ 
+2\delta_1&+2\delta_1+4\delta_2+2\delta_3&-3\delta_1-2\delta_2-\delta_3&0\\
}.
\]
Here the vertical bar separates the components $A_{\ast,0}$ which couple 
to the length $L$ 
from the components $A_{\ast,j}$ which couple to $K_{j}$. 
The horizontal bar separates the components $A_{0,\ast}$ for
the momentum constraint from the components $A_{j,\ast}$ which couple 
to the Bethe equation for $u_{j,k}$. 
In other words, the deformation of the Bethe equations 
of \cite{Minahan:2002ve} is
\begin{eqnarray}
1\eq
e^{-2i\delta_3L}
e^{i(+\delta_1+2\delta_2+3\delta_3)K_2}
e^{i(-2\delta_1-4\delta_2-2\delta_3)K_3}
\nl\times
\mathord{\phantom{\left(\frac{u_{1,k}-\ialf}{u_{1,k}+\ialf}\right)^L}}\,
\mathord{\phantom{\prod_{j=1}^{K_2}
\frac{u_{1,k}-u_{2,j}-\ialf}{u_{1,k}-u_{2,j}+\ialf}}}\,
\prod_{\textstyle\atopfrac{j=1}{j\neq k}}^{K_1}
\frac{u_{1,k}-u_{1,j}+i}{u_{1,k}-u_{1,j}-i}\,
\prod_{j=1}^{K_2}
\frac{u_{1,k}-u_{2,j}-\ialf}{u_{1,k}-u_{2,j}+\ialf}\,,
\cr
1\eq
e^{i(-\delta_1+\delta_3)L}
e^{i(-\delta_1-2\delta_2-3\delta_3)K_1}
e^{i(+3\delta_1+2\delta_2+\delta_3)K_3}
\nl\times
\left(\frac{u_{2,k}-\ialf}{u_{2,k}+\ialf}\right)^L\,
\prod_{j=1}^{K_1}
\frac{u_{2,k}-u_{1,j}-\ialf}{u_{2,k}-u_{1,j}+\ialf}
\prod_{\textstyle\atopfrac{j=1}{j\neq k}}^{K_2}
\frac{u_{2,k}-u_{2,j}+i}{u_{2,k}-u_{2,j}-i}
\prod_{j=1}^{K_3}
\frac{u_{2,k}-u_{3,j}-\ialf}{u_{2,k}-u_{3,j}+\ialf}\,,
\cr
1\eq
e^{+2i\delta_1L}
e^{(+2\delta_1+4\delta_2+2\delta_3)K_1}
e^{(-3\delta_1-2\delta_2-\delta_3)K_2}
\nl\times
\mathord{\phantom{\left(\frac{u_{1,k}-\ialf}{u_{1,k}+\ialf}\right)^L}}\,
\prod_{j=1}^{K_2}
\frac{u_{3,k}-u_{2,j}-\ialf}{u_{3,k}-u_{2,j}+\ialf}
\prod_{\textstyle\atopfrac{j=1}{j\neq k}}^{K_3}
\frac{u_{3,k}-u_{3,j}+i}{u_{3,k}-u_{3,j}-i}\,,
\end{eqnarray}
and the cyclicity constraint becomes
\[
1=e^{+2i\delta_3K_1}e^{i(+\delta_1-\delta_3)K_2}e^{-2i\delta_1K_3}
\prod_{j=1}^{K_2}\frac{u_{2,j}+\ialf}{u_{2,j}-\ialf}\,.
\]
%

\subsection{Fermions and compatibility with Feynman diagrams}
\label{eq:Compatibility}

Here we will generalize the results of the previous section to
the full $\superN=4$ parent theory. 
In addition to the three scalars $\phi_j$ transforming canonically
under some $\alg{su}(3)$ there are three flavored fermions
$\psi_j$ transforming in the same representation
and one $\alg{su}(3)$-invariant gluino $\chi$.
Let us for the moment generalize the deformation 
and make it depend not on the charges of the fields, 
but introduce an independent phase for any set of interacting fields.
We deform the couplings of the scalars and the fermions
by the phases $\alpha_j,\beta_j,\gamma_j,\alpha'_j,\beta'_j$
(the indices of fields are identified modulo $\IZ_3$)%
\footnote{The phases $\alpha'_j,\beta'_j$ play 
no role in the diagonal scattering terms
investigated below, but for integrability
they are necessarily zero, $\alpha'_j=\beta'_j=0$,
due to off-diagonal scattering.}
\[e^{i\alpha'_j}\Tr \phi_j\comm{\psi_{j+1}}{\psi_{j-1}}_{\alpha_j},\qquad
e^{i\beta'_j}\Tr \bar\phi^j\comm{\chi}{\psi_j}_{\beta_j},\qquad
\Tr\comm{\phi_{j-1}}{\phi_{j+1}}_{\gamma_j}\comm{\bar\phi^{j-1}}{\bar\phi^{j+1}}_{\gamma_j}.
\]
with the twisted commutator $\comm{X}{Y}_\alpha=e^{+i\alpha/2} XY-e^{-i\alpha/2} YX$.
These are the most general renormalizable deformations of the $\superN=4$ model
using only phases.
We now collect the fields in a vector
$(\phi_1,\phi_2,\phi_3\mathpunct{|}\psi_1,\psi_2,\psi_3,\chi)$
and determine the twist matrix by combining the 
Yukawa vertices appropriately
\[\label{eq:GenPhases}
\mat{B}=\matr{ccc|cccc}{
        0&-\gamma_3&+\gamma_2& +\beta_1&+\alpha_1&-\alpha_1& -\beta_1 \\
+\gamma_3&        0&-\gamma_1&-\alpha_2& +\beta_2&+\alpha_2& -\beta_2 \\
-\gamma_2&+\gamma_1&        0&+\alpha_3&-\alpha_3& +\beta_3& -\beta_3 \\\hline
 -\beta_1&+\alpha_2&-\alpha_3&        0&+\alpha_3&-\alpha_2& -\beta_1 \\
-\alpha_1& -\beta_2&+\alpha_3&-\alpha_3&        0&+\alpha_1& -\beta_2 \\
+\alpha_1&-\alpha_2& -\beta_3&+\alpha_2&-\alpha_1&        0& -\beta_3 \\
 +\beta_1& +\beta_2& +\beta_3& +\beta_1& +\beta_2& +\beta_3&        0
}.
\]

\begin{figure}\centering
\begin{minipage}{260pt}
\setlength{\unitlength}{1pt}%
\small\thicklines%
\begin{picture}(260,20)(-10,-10)
\put(  0,00){\circle{15}}%
\put(  7,00){\line(1,0){26}}%
\put( 40,00){\circle{15}}%
\put( 47,00){\line(1,0){26}}%
\put( 80,00){\circle{15}}%
\put( 87,00){\line(1,0){26}}%
\put(120,00){\circle{15}}%
\put(127,00){\line(1,0){26}}%
\put(160,00){\circle{15}}%
\put(167,00){\line(1,0){26}}%
\put(200,00){\circle{15}}%
\put(207,00){\line(1,0){26}}%
\put(240,00){\circle{15}}%
\put( 35,-5){\line(1, 1){10}}%
\put( 35, 5){\line(1,-1){10}}%
\put(195,-5){\line(1, 1){10}}%
\put(195, 5){\line(1,-1){10}}%
\put( 00,00){\makebox(0,0){$-$}}%
\put( 80,00){\makebox(0,0){$+$}}%
\put(120,00){\makebox(0,0){$+$}}%
\put(160,00){\makebox(0,0){$+$}}%
\put(240,00){\makebox(0,0){$-$}}%
\end{picture}
\end{minipage}

\caption{``Beauty'{}' Dynkin diagram of $\alg{su}(2,2|4)$.}
\label{fig:DynkinBeauty}
\end{figure}

{}From a field theory point of view, arbitrary values of 
$\alpha_j,\beta_j,\gamma_j$ are allowed. 
However, not all values necessarily preserve integrability. 
To determine the relations among the phases we consider
the set of excitations used in the Bethe equations. 
For the one-loop Bethe equations \cite{Beisert:2003yb},
there are seven types of excitations $\mathcal{B}_j$. These depend on
the Cartan matrix which is not uniquely determined 
for a superalgebra. Here we use the Cartan matrix 
of $\alg{su}(2,2|4)$ corresponding to the ``Beauty{}'' 
Dynkin diagram \cite{Beisert:2003yb} in \figref{fig:DynkinBeauty}.
Then the excitations $\mathcal{B}_1,\mathcal{B}_7$ merely act on 
the spacetime part of algebra and there are no deformations. 
All the other excitations involve flavor degrees of freedom%
\footnote{The derivative was included for completeness. For the purposes
of the current investigation, it could dropped as it does not
carry flavor charges.}
\[\label{eq:BeautyEx}
\begin{array}[b]{cllllll}
\mathcal{B}_2:& \bar\phi^3\mapsto\bar\psi^3, & \bar\phi^2\mapsto\bar\psi^2, & \bar\phi^1\mapsto\bar\psi^1,
    & \psi_3\mapsto D\phi_3, & \psi_2\mapsto D\phi_2, & \psi_1\mapsto D\phi_1,
\\[0.3em]
\mathcal{B}_3:& \phi_2\mapsto\bar\phi^3, & \phi_3\mapsto\bar\phi^2, & \chi\mapsto\psi_1, & \bar\psi^1\mapsto\bar\chi,
\\[0.3em]
\mathcal{B}_4:& \phi_1\mapsto\phi_2, & \bar\phi^2\mapsto\bar\phi^1, & \psi_1\mapsto\psi_2, & \bar\psi^2\mapsto\bar\psi^1,
\\[0.3em]
\mathcal{B}_5:& \phi_2\mapsto\phi_3, & \bar\phi^3\mapsto\bar\phi^2, & \psi_2\mapsto\psi_3, & \bar\psi^3\mapsto\bar\psi^2,
\\[0.3em]
\mathcal{B}_6:& \phi_3\mapsto\chi, & \bar\phi^2\mapsto\psi_1, & \bar\phi^1\mapsto\psi_2,
    & \bar\chi\mapsto D\bar\phi^3, & \bar\psi^1\mapsto D\phi_2, & \bar\psi^2\mapsto D\phi_1.
\end{array}
\]
When commuting two such excitations we should expect a definite phase
in order for integrability to be preserved.
However, the excitations can act on various types of fields and depending 
on the type, there will be a different phase shift 
when we rely on the most general form \eqref{eq:GenPhases}.
We therefore have to impose certain constraints 
on the $\alpha_j,\beta_j,\gamma_j$, namely
that the combination of excitations $\mathcal{B}^{}_j\mathcal{B}^{-1}_j$
must be trivial. This should remain true even when choosing 
two different actions for $\mathcal{B}^{}_j$ and $\mathcal{B}^{-1}_j$ 
out of the various allowed ones.
We can rewrite $\mathcal{B}^{}_j\mathcal{B}^{-1}_j$
as a combination of four fields which should not yield a phase
when commuting past anything; we find the following three independent ones
\[
\psi_1\phi_2\phi_3\bar\chi\hateq\psi_2\phi_3\phi_1\bar\chi\hateq\psi_3\phi_1\phi_2\bar\chi\hateq1.
\]
For the purpose of determining the phases, they are as good as any number.
They correspond to the following vectors in the space of fields
\<
\vect{v}_1\eq(0,1,1\mathpunct{|}1,0,0,-1),\nln
\vect{v}_2\eq(1,0,1\mathpunct{|}0,1,0,-1),\nln
\vect{v}_3\eq(1,1,0\mathpunct{|}0,0,1,-1).
\>
The consistency condition, i.e.~the absence of twists for these combinations, 
now implies
\[
\mat{B}\vect{v}_1=\mat{B} \vect{v}_2=\mat{B} \vect{v}_3=0.
\]
This leads to the following relations among the phases
\[\label{eq:PhasePara}
\begin{array}{rclcrclcrcl}
\alpha_1\eq+\delta_1,&\quad&
\beta_1\eq+\delta_3,&\quad&
\gamma_1\eq-\delta_1-2\delta_2-\delta_3,
\\[0.3em]
\alpha_2\eq+\delta_1+\delta_2,&&
\beta_2\eq-\delta_2-\delta_3,&&
\gamma_2\eq-\delta_1-\delta_3,
\\[0.3em]
\alpha_3\eq+\delta_1+\delta_2+\delta_3,&&
\beta_3\eq+\delta_2, &&
\gamma_3\eq-\delta_1+\delta_3,
\end{array}
\]
which we have parametrized using the phases $\delta_j$.
The twist matrix $\mat{B}$ with phases \eqref{eq:PhasePara} 
therefore leads to an integrable system.

We can now transform the twist matrix $\mat{B}$ to the form 
required by the Bethe equations. 
The new basis is 
$(\phi_1\mathpunct{|}\mathcal{B}_1,\mathcal{B}_2,\mathcal{B}_3,\mathcal{B}_4,\mathcal{B}_5,\mathcal{B}_6,\mathcal{B}_7)$
and, when 
expressed in terms of fields, it is equivalent to
$(\phi_1\mathpunct{|}1,\phi_1\phi_2\phi_3\bar\chi,\bar\phi^2\bar\phi^3,
\bar\phi^1\phi_2,\bar\phi^2\phi_3,\bar\phi^3\chi,1)$.
{}From \eqref{eq:GenPhases,eq:PhasePara}
we read off the twist matrix for the Bethe equations
\[\label{eq:TwistBeauty}
\mat{A}=
\left(
\mbox{\scriptsize$\displaystyle
\begin{array}{c|ccccccc}
0&0&-\delta_3&+2\delta_3&+\delta_1-\delta_3&-2\delta_1&+\delta_1&0\\\hline
0&0&0&0&0&0&0&0\\
+\delta_3&0&0&+\delta_3&-\delta_2-2\delta_3&+2\delta_2+\delta_3&-\delta_2&0\\
-2\delta_3&0&-\delta_3&0&+\delta_1+2\delta_2+3\delta_3&-2\delta_1-4\delta_2-2\delta_3&+\delta_1+2\delta_2&0\\ 
-\delta_1+\delta_3&0&+\delta_2+2\delta_3&-\delta_1-2\delta_2-3\delta_3&0&+3\delta_1+2\delta_2+\delta_3&-2\delta_1-\delta_2&0\\ 
+2\delta_1&0&-2\delta_2-\delta_3&+2\delta_1+4\delta_2+2\delta_3&-3\delta_1-2\delta_2-\delta_3&0&+\delta_1&0\\
-\delta_1&0&+\delta_2&-\delta_1-2\delta_2&+2\delta_1+\delta_2&-\delta_1&0&0\\
0&0&0&0&0&0&0&0
\end{array}$}
\right).
\]
As can be seen, the middle $3\times 3$ block together with the first
row/column agrees with the matrix \eqref{eq:TwistSO6}. Also, the lower
right $4\times 4$ block together with the first
row/column agrees, up to the change of variables \eqref{eq:PhasePara},
with the matrix \eqref{eq:Twistu23}. It therefore follows that the only
flavor twist compatible with the Feynman rules is the one discussed in
\Secref{sec:ChTw}.

The equation 
\eqref{eq:TwistBeauty} implies that, for certain choices of twist
parameters, nontrivial subalgebras of the superconformal algebra
survive in the deformed theory. For $\delta_2=\delta_3=0$, the third
row/column disappears. 
We can then add roots of flavor $2$ at infinity $u_2=\infty$ to the
set of Bethe roots without spoiling the equations or changing the
energy.
This means that not only conformal symmetry $\alg{su}(2,2)$ is preserved, 
but also supersymmetry, i.e.~$\superN=1$ superconformal $\alg{su}(2,2|1)$
symmetry. 
Similarly, we can set $\delta_1=\delta_2=0$ to preserve (a different) 
$\superN=1$ supersymmetry.
There are three other apparent choices of phases for which some
of the symmetry is preserved: $\delta_1=2\delta_2+\delta_3=0$,
$2\delta_1+\delta_2=\delta_2+2\delta_3=0$ or
$\delta_1+2\delta_2=\delta_3=0$.
In these cases one of the three central rows/columns vanishes 
and a $\alg{su}(2)$ factor of the internal $\alg{su}(4)$ symmetry 
survives. Let us summarize the conditions for preserved symmetries
\[\label{eq:SymCharges}
\begin{array}{rc}
\superN=1\colon & \delta_2=\delta_3=0,
\\[0.3em]
\alg{su}(2)\colon & \delta_1=2\delta_2+\delta_3=0,
\\[0.3em]
\alg{su}(2)\colon & 2\delta_1+\delta_2=\delta_2+2\delta_3=0,
\\[0.3em]
\alg{su}(2)\colon & \delta_1+2\delta_2=\delta_3=0,
\\[0.3em]
\superN=1\colon & \delta_1=\delta_2=0.
\end{array}
\]

To improve our understanding of twist matrices, let us see
how to obtain $\mat{A}$ more directly. We introduce three charge vectors
\<\label{eq:ChargeDynkin}
\vect{q}_{q_1}\eq(\phantom{+}0\mathpunct{|}\phantom{+}0,+1,-2,+1,\phantom{+}0,\phantom{+}0,\phantom{+}0),\nln
\vect{q}_{p_{\hphantom{1}}}  \eq(+1\mathpunct{|}\phantom{+}0,\phantom{+}0,+1,-2,+1,\phantom{+}0,\phantom{+}0),\nln
\vect{q}_{q_2}\eq(\phantom{+}0\mathpunct{|}\phantom{+}0,\phantom{+}0,\phantom{+}0,+1,-2,+1,\phantom{+}0).
\>
These can be used to extract the Dynkin labels $[q_1,p,q_2]$ of a state 
from the number of excitations
$\vect{K}=(L\mathpunct{|}K_1,K_2,K_3,K_4,K_5,K_6,K_7)$
by, cf.~\cite{Beisert:2003yb}
\[
\vect{q}_{q_1}\cdott\vect{K}=q_1,\qquad
\vect{q}_{p}\cdott\vect{K}=p,\qquad
\vect{q}_{q_2}\cdott\vect{K}=q_2.
\]
The matrix $\mat{A}$ is then given as follows
\[\label{eq:TwistDynkin}
\mat{A}=
\delta_1
\lrbrk{\vect{q}_{p}^{}\vect{q}_{q_2}^\trans
-\vect{q}_{q_2}^{}\vect{q}_{p}^\trans}
+\delta_2
\lrbrk{\vect{q}_{q_2}^{}\vect{q}_{q_1}^\trans
-\vect{q}_{q_1}^{}\vect{q}_{q_2}^\trans}
+\delta_3
\lrbrk{\vect{q}_{q_1}^{}\vect{q}_{p}^\trans
-\vect{q}_{p}^{}\vect{q}_{q_1}^\trans}.
\]
%

\subsection{Dualizations}
\label{sec:Dualize}

For a superalgebra there are various equivalent 
forms of the Bethe equations which correspond to
various equivalent Dynkin diagrams. 
The Bethe roots for one state 
in the various forms 
are related by dualization. 
A dualization replaces all Bethe roots
of one fermionic type by dualized roots
while preserving all the other roots. 
The new set of roots obeys dualized
Bethe equations corresponding to the
dualized Dynkin diagram. 
The procedure of dualization
was found in \cite{Essler:1992nk,Essler:1992uc}
and is described in detail in \cite{Beisert:2005di}
for the current setup.

We now follow the steps in \cite{Beisert:2005di}
for a dualization of flavor $j$
and trace the insertions of phases. 
We see that the phases from the equation for flavor $j$ 
will be added to the phases in the equations for flavors 
$j\pm 1$. 
Afterwards the equation for flavor $j$ is inverted.
On the twist matrix this has the following effect
on the rows
\[\label{eq:DualRows}
\matr{c}{\vdots\\A_{j-1,\ast}\\A_{j,\ast}\\A_{j+1,\ast}\\\vdots}
\to
\matr{c}{\vdots\\A_{j-1,\ast}+A_{j,\ast}\\-A_{j,\ast}\\A_{j+1,\ast}+A_{j,\ast}\\\vdots}.
\]
There is however another effect which has to be taken into account:
the twist matrix is multiplied to the vector of 
excitations which changes according to%
\footnote{Here we cannot subtract $1$ as in \cite{Beisert:2005di} 
(corresponding to a root at $\infty$ which we conventionally remove in
the undeformed theory),
because the leading terms in the polynomial $P(u)$ do not cancel.
This is because the breaking of the symmetry of the parent theory.
The states in a parent multiplet no longer have a common energy.}
\[
K_j\to K_{j+1}+K_{j-1}-K_j.
\]
This change is absorbed by the further transformation
on columns
\[\label{eq:DualColumns}
\matr{ccccc}{\ldots&A_{\ast,j-1}&A_{\ast,j}&A_{\ast,j+1}&\ldots}
\to
\matr{ccccc}{\ldots&A_{\ast,j-1}+A_{\ast,j}&-A_{\ast,j}&A_{\ast,j+1}+A_{\ast,j}&\ldots}.
\]
It is the same transformation as
\eqref{eq:DualRows} and thus antisymmetry of $\mat{A}$ is preserved.

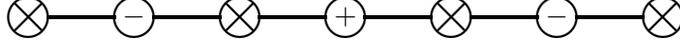
\begin{figure}\centering
\begin{minipage}{260pt}
\setlength{\unitlength}{1pt}%
\small\thicklines%
\begin{picture}(260,20)(-10,-10)
\put(  0,00){\circle{15}}%
\put(  7,00){\line(1,0){26}}%
\put( 40,00){\circle{15}}%
\put( 47,00){\line(1,0){26}}%
\put( 80,00){\circle{15}}%
\put( 87,00){\line(1,0){26}}%
\put(120,00){\circle{15}}%
\put(127,00){\line(1,0){26}}%
\put(160,00){\circle{15}}%
\put(167,00){\line(1,0){26}}%
\put(200,00){\circle{15}}%
\put(207,00){\line(1,0){26}}%
\put(240,00){\circle{15}}%
\put( -5,-5){\line(1, 1){10}}%
\put( -5, 5){\line(1,-1){10}}%
\put( 75,-5){\line(1, 1){10}}%
\put( 75, 5){\line(1,-1){10}}%
\put(155,-5){\line(1, 1){10}}%
\put(155, 5){\line(1,-1){10}}%
\put(235,-5){\line(1, 1){10}}%
\put(235, 5){\line(1,-1){10}}%
\put( 40,00){\makebox(0,0){$-$}}%
\put(120,00){\makebox(0,0){$+$}}%
\put(200,00){\makebox(0,0){$-$}}%
\end{picture}
\end{minipage}

\caption{Dynkin diagram of $\alg{su}(2,2|4)$
for the higher-loop Bethe equations.}
\label{fig:DynkinHigher}
\end{figure}

For example, we can now transform the twist matrix
to the Dynkin diagram 
for the higher-loop Bethe equations
used in \cite{Beisert:2005fw} 
with $\eta_1=\eta_2=+1$, cf.~\figref{fig:DynkinHigher}.
This is done most conveniently by preserving the
form \eqref{eq:TwistDynkin} and merely transforming
the charge vectors \eqref{eq:ChargeDynkin}
according to \eqref{eq:DualColumns}.
The new charge vectors are
\<\label{eq:ChargeDynkinDual}
\vect{q}_{q_1}\eq(\phantom{+}0\mathpunct{|}-1,\phantom{+}0,-1,+1,\phantom{+}0,\phantom{+}0,\phantom{+}0),\nln
\vect{q}_{p_{\hphantom{1}}}  \eq(+1\mathpunct{|}\phantom{+}0,\phantom{+}0,+1,-2,+1,\phantom{+}0,\phantom{+}0),\nln
\vect{q}_{q_2}\eq(\phantom{+}0\mathpunct{|}\phantom{+}0,\phantom{+}0,\phantom{+}0,+1,-1,\phantom{+}0,-1)
\>
and the twist matrix is
\[\label{eq:TwistDual}
\mat{A}=
\left(
\mbox{\scriptsize$\displaystyle
\begin{array}{c|ccccccc}
0&+\delta_3&0&+\delta_3&+\delta_1-\delta_3&-\delta_1&0&-\delta_1\\\hline
-\delta_3&0&0&-\delta_3&+\delta_2+2\delta_3&-\delta_2-\delta_3&0&-\delta_2\\
0&0&0&0&0&0&0&0\\
-\delta_3&+\delta_3&0&0&+\delta_1+\delta_2+\delta_3&-\delta_1-\delta_2-\delta_3&0&-\delta_1-\delta_2\\ 
-\delta_1+\delta_3&-\delta_2-2\delta_3&0&-\delta_1-\delta_2-\delta_3&0&+\delta_1+\delta_2+\delta_3&0&+2\delta_1+\delta_2\\ 
+\delta_1&+\delta_2+\delta_3&0&+\delta_1+\delta_2+\delta_3&-\delta_1-\delta_2-\delta_3&0&0&-\delta_1\\
0&0&0&0&0&0&0&0\\
+\delta_1&+\delta_2&0&+\delta_1+\delta_2&-2\delta_1-\delta_2&\delta_1&0&0
\end{array}$}
\right).
\]
In this form one row/column vanishes when we set
\[\label{eq:SymChargesDual}
\begin{array}{rc}
\superN=1\colon & \delta_2=\delta_3=0,
\\[0.3em]
\superN=1\colon & \delta_1+\delta_2=\delta_3=0,
\\[0.3em]
\alg{su}(2)\colon & 2\delta_1+\delta_2=\delta_2+2\delta_3=0,
\\[0.3em]
\superN=1\colon & \delta_1=\delta_2+\delta_3=0=0,
\\[0.3em]
\superN=1\colon & \delta_1=\delta_2=0.
\end{array}
\]
Note that as compared to \eqref{eq:SymCharges}
two of the conditions 
have been replaced by different ones which now
preserve $\superN=1$ supersymmetry instead of
$\alg{su}(2)$.
As both forms of the Bethe equations are equivalent, 
all the relations between the angles in
\eqref{eq:SymCharges,eq:SymChargesDual}
must lead to symmetries. Some of the symmetries
are however hidden in one form and manifest in the other.

\subsection{Higher loops}
\label{sec:Higher}

It has turned out that the planar dilatation operator of 
$\superN=4$ SYM is not only integrable
at one loop, but also at higher loops \cite{Beisert:2003tq},
i.e.~higher order in the coupling constant
\[
g^2=\frac{\lambda}{8\pi^2}\,.
\]
In \cite{Beisert:2005fw} Bethe equations for this long-range chain 
have been proposed. These are a generalization of the Bethe
equations for the $\alg{su}(2)$ sector 
at higher loops \cite{Serban:2004jf,Beisert:2004hm}
which are somewhat similar to the ones 
of the Inozemtsev spin chain \cite{Inozemtsev:1989yq,Inozemtsev:2002vb}.
Let us briefly summarize the
equations and refer the reader to \cite{Beisert:2005fw} for all
details. The most convenient way of parametrizing the Bethe equations
is to treat the rapidities $u_{j,k}$ as composite quantities
\cite{Beisert:2004jw} related to the more fundamental variables $x_{j,k}$
by
\[
u_{j,k}=u(x_{j,k}),
\qquad
u(x)=x+g^2/2x.
\]
Similarly, we define the derived quantity $x^\pm_{j,k}$ as 
\[
x^\pm_{j,k}=x(u_{j,k}\pm \ihalf ),
\qquad
x(u)=\half u+\half u\sqrt{1-2g^2/u^2}\,.
\]
The higher-loop Bethe equations and momentum constraint 
proposed in \cite{Beisert:2005fw} read
\[\label{eq:BetheHigher}
U_0=1,\qquad
U_j(x_{j,k})
\mathop{\prod_{j'=1}^7\prod_{k'=1}^{K_{j'}}}_{(j',k')\neq(j,k)}
\frac{u_{j,k}-u_{j',k'}+\ihalf M_{j,j'}}{u_{j,k}-u_{j',k'}-\ihalf M_{j,j'}}
=1,
\]
with $M_{j,j'}$ the Cartan matrix specified by \figref{fig:DynkinHigher}.
The deformation of the (untwisted) one-loop equations 
\eqref{eq:TwistBethe,eq:TwistMomentum}
is contained in the terms $U_j$ which read
\[
U_0=
\prod_{k=1}^{K_4}
\frac{x^+_{4,k}}{x^-_{4,k}}\,,
\qquad
U_1(x)=U_3^{-1}(x)=U_5^{-1}(x)=U_7(x)=
\prod_{k=1}^{K_4}
S\indup{aux}(x_{4,k},x)
\]
as well as
\[
U_4(x)=
U\indup{s}(x)
\lrbrk{\frac{x^-}{x^+}}^L
\prod_{k=1}^{K_1}
S^{-1}\indup{aux}(x,x_{1,k})
\prod_{k=1}^{K_3}
S\indup{aux}(x,x_{3,k})
\prod_{k=1}^{K_5}
S\indup{aux}(x,x_{5,k})
\prod_{k=1}^{K_7}
S^{-1}\indup{aux}(x,x_{7,k}).
\]
The auxiliary scattering term $S\indup{aux}(x_1,x_2)$ is defined as
\[
S\indup{aux}(x_1,x_2)=
\frac{1-g^2/2x^+_1x_2}{1-g^2/2x^-_1x_2}\,.
\]
The dressing factor $U\indup{s}(x)$ is some function of the $x_{4,k}$
which can be used to deform the model.
For gauge theory it is trivial, $U\indup{s}(x)=1$.
The anomalous dimension of a state is given by
\[
\delta D=g^2\sum_{k=1}^{K_4}\lrbrk{\frac{i}{x^+_{4,k}}-\frac{i}{x^-_{4,k}}}.
\]

A complication of the higher-loop model is that 
the length of the chain is not preserved by the Hamiltonian,
the spin chain becomes \emph{dynamic} at higher loops
\cite{Beisert:2003ys}.
It was argued that the proposed Bethe equations 
have a symmetry which enables one to interpret a set
of Bethe roots as a spin chain state with flexible length.
This \emph{dynamic transformation} is a prescription 
of how to change a set of Bethe roots for length $L$ 
into a set of Bethe roots for length $L+1$. 
It replaces a root of type $3$ by a root of type $1$
\[
x_3\to x_1=g^2/2x_3
\]
so that the excitation numbers change according to
\[\label{eq:Dynamic1}
K_3\to K_3-1,\quad
K_1\to K_1+1,\quad
L\to L+1,\quad
B\to B-1.
\]
Here, $B$ is the hypercharge of the state.
The transformation is a symmetry of the Bethe equations
due to the identities
\[
u(x_3)=u(g^2/2x_3),\qquad
\frac{x_{4,k}^+}{x_{4,k}^-}\,
S\indup{aux}(x_{4,k},x_3)
S\indup{aux}(x_{4,k},g^2/2x_3)
=
\frac{u_{4,k}-u_3+\ihalf }{u_{4,k}-u_3-\ihalf }\,.
\]
This proves the invariance of the Bethe equation for $x_{4,k}$ and 
leads to the identity
\[
U_3(x_3)
\prod_{k=1}^{K_4}\frac{u_3-u_{4,k}-\ihalf }{u_3-u_{4,k}+\ihalf }
=U_0\,U_1(g^2/2x_3)
\]
which relates the Bethe equations of $x_3$ and $x_1=g^2/x_3$.
Note that $M_{3,4}=-1$ and $M_{1,4}=0$.
An analogous transformation replaces a root of type $5$ by 
a root of type $7$
\[\label{eq:Dynamic2}
K_5\to K_5-1,\quad
K_7\to K_7+1,\quad
L\to L+1,\quad
B\to B+1.
\]

We would now like to introduce the twist into the higher-loop model.
The twisting of the spin chain is a discrete procedure:
there is a fixed phase for an interchange of a pair of fields.
Moreover, the phase depends only on the conserved charges of the fields.
This implies that the twisting is independent of the loop order. 
In particular it can also be inferred from the position space
Bethe ansatz in \cite{Staudacher:2004tk,Beisert:2005fw}.
Let us therefore twist the higher-loop Bethe equations in the most obvious way, 
we multiply the Bethe equations and momentum constraint \eqref{eq:BetheHigher}
by the appropriate phases as in \eqref{eq:TwistBethe,eq:TwistMomentum}
\[\label{eq:TwistBetheHigher}
e^{i(\mat{A}\vect{K})_0}
\,U_0
=1,\qquad
e^{i(\mat{A}\vect{K})_j}\,
U_j(x_{j,k})
\mathop{\prod_{j'=1}^7\prod_{k'=1}^{K_{j'}}}_{(j',k')\neq(j,k)}
\frac{u_{j,k}-u_{j',k'}+\ihalf M_{j,j'}}{u_{j,k}-u_{j',k'}-\ihalf M_{j,j'}}
=1.
\]
This deformation is consistent with the twist \cite{Frolov:2005ty} 
of the higher-loop Bethe equations in
the $\alg{su}(2)$ sector \cite{Arutyunov:2004vx}.

Does this interfere with the dynamic transformation?
Let us first consider the Bethe equations for those roots
which are not transformed. 
According to \eqref{eq:Dynamic1} we increase each of 
$L,K_1,-K_3$ by one unit. As it turns out, 
the sum of the first and second column in \eqref{eq:TwistDual}
equals the fourth column. Therefore these Bethe equations
are not modified by the transformation.
We need to confirm that also the Bethe equation for the transformed root
does not change. The total twist in Bethe equations for roots
of type $1$ and type $3$ is different. 
However, even the untwisted equations are not immediately the same,
but only after making use of the momentum constraint.
Now the momentum constraint is twisted as well, 
cf. \eqref{eq:TwistMomentum},
and it accounts for the difference of phases. 
In \eqref{eq:TwistDual} we can see
that the first row (which enters the momentum constraint) 
plus the third row equals the fourth row.
In fact, this is obvious due to the antisymmetry of $\mat{A}$
and the above observations.
A similar argument holds for the other dynamic transformation 
\eqref{eq:Dynamic2}.
Twisting of the proposed higher-loop Bethe equations 
is therefore consistent. 

The deeper reason why the twist is possible is that it only
depends on the Dynkin labels $[q_1,p,q_2]$ of the state and not on
on the length $L$ and the hypercharge $B$.
The Dynkin labels are physical and they are invariant under
the dynamic transformation.
Put differently, the combination which enters the
Bethe equations, $\mat{A}\vect{K}$, depends
on $[q_1,p,q_2]$ only, but not on $L,B$.
Curiously, we did not explicitly use this constraint
when deriving \eqref{eq:TwistDual}. 
Instead we demanded compatibility of 
Feynman rules with integrability. 
This however implicitly leads to the independence of $L$ and $B$
because the diagrams do not conserve these quantities.

\section{Deformations Involving Spacetime}
\label{sec:NonCom}

In \secref{sec:mostgeneral} 
we have investigated the most general twisting
on a standard integrable spin chain. 
We have seen that it is specified by a generic antisymmetric
matrix $\mat{A}$ with 28 free parameters.
In \secref{eq:Compatibility} we have then seen that
not every such twisted standard integrable spin chain can be 
realized at one loop by a twist of $\superN=4$ SYM. 
Conversely, not every twist of $\superN=4$ SYM has a one loop 
dilatation operator which is integrable.
Here we would like to investigate the set of all integrable twists of 
$\superN=4$ SYM.

\subsection{Bethe ansatz}
\label{sec:NCBethe}

The starting point will be the independence of
$\mat{A}\vect{K}$ of the length $L$ and 
the hypercharge $B$. In section \secref{eq:Compatibility} this was a
consequence of the requirement of compatibility of Feynman
diagrammatics and integrability. As we have seen in
\secref{sec:Higher}, this independence is necessary for consistency of
the higher-loop Bethe equations. We will therefore require it in the
following.

To construct a matrix $\mat{A}$, we shall use a form similar to
\eqref{eq:TwistDynkin}
\[\label{eq:TwistDynkin2}
\mat{A}=
\sum_{j,j'}
\alpha_{j,j'}
\lrbrk{\vect{q}_{j}^{}\vect{q}_{j'}^\trans
-\vect{q}_{j'}^{}\vect{q}_{j}^\trans}~,
\]
where the $\vect{q}_{j}$ are the allowed charge vectors.
We have used the charge vectors \eqref{eq:ChargeDynkinDual2}
\<\label{eq:ChargeDynkinDual2}
\vect{q}_{q_1}\eq(\phantom{+}0\mathpunct{|}-1,\phantom{+}0,-1,+1,\phantom{+}0,\phantom{+}0,\phantom{+}0),\nln
\vect{q}_{p}  \eq(+1\mathpunct{|}\phantom{+}0,\phantom{+}0,+1,-2,+1,\phantom{+}0,\phantom{+}0),\nln
\vect{q}_{q_2}\eq(\phantom{+}0\mathpunct{|}\phantom{+}0,\phantom{+}0,\phantom{+}0,+1,-1,\phantom{+}0,-1)
\>
dual to the Dynkin labels $[q_1,p,q_2]$ of $\alg{su}(4)$.
Similarly, we could use the charge vectors 
\<\label{eq:ChargeDynkinDualAdS}
\vect{q}_{s_1}\eq(\phantom{+}0\mathpunct{|}-2,+1,\phantom{+}0,\phantom{+}0,\phantom{+}0,\phantom{+}0,\phantom{+}0),\nln
\vect{q}_{r_0}\eq(-1\mathpunct{|}+1,-1,\phantom{+}0,\phantom{+}0,\phantom{+}0,-1,+1),\nln
\vect{q}_{s_2}\eq(\phantom{+}0\mathpunct{|}\phantom{+}0,\phantom{+}0,\phantom{+}0,\phantom{+}0,\phantom{+}0,+1,-2),
\>
which are dual to the Dynkin labels $[s_1,r,s_2]$ of the conformal
algebra $\alg{su}(2,2)$. The labels $s_1,s_2$ are twice the Lorentz spins
and $r=-D-\half s_1-\half s_2$ with $D$ the scaling dimension.
To complete the basis of the eight-dimensional space we introduce
\<\label{eq:ChargeDynkinDualUnphys}
\vect{q}_{L}\eq(+1\mathpunct{|}\phantom{+}0,\phantom{+}0,\phantom{+}0,\phantom{+}0,\phantom{+}0,\phantom{+}0,\phantom{+}0),\nln
\vect{q}_{B}\eq(\phantom{+}0\mathpunct{|}-\half,\phantom{+}0,+\half,\phantom{+}0,-\half,\phantom{+}0,+\half),
\>
which are dual to $L$ and $B$. 
Clearly, the latter two are not suitable
for the construction of $\mat{A}$ in \eqref{eq:TwistDynkin2}.%
\footnote{They can be used for the one-loop Bethe equations
but these equations do not correspond to a deformation 
of $\superN=4$ SYM. The higher-loop Bethe equations 
are incompatible with $\vect{q}_{L}$ and $\vect{q}_{B}$.}

It is questionable whether we should use $\vect{q}_{r_0}$ or not.
All the other labels $q_1,p,q_2,s_1,s_2$ are positive integers.
Conversely, $r=-D_0-\half s_1-\half s_2-\delta D$ is irrational
in almost all cases due to the anomalous dimension $\delta D$. 
However, in the way we have twisted the equations, 
$\vect{q}_{r_0}$ does not actually couple to $r$, 
but to its classical limit 
\[
\vect{q}_{r_0}\cdott\vect{K}=r_0=-D_0-\half s_1-\half s_2.
\]
We could also allow for twisting using the exact label $r$ or,
equivalently, using the anomalous dimension $\delta D$.
{}From a spin chain point of view, 
this may be possible and will yield deformations 
(see \cite{Beisert:2005fw} for notation)
\[
\exp\bigbrk{ig^2 Q_2(g)}
\quad\mbox{or}\quad
\exp\bigbrk{ig^2 K_j q_{2}(x_{4,k})},
\]
which are somewhat similar to the 
deformation discussed in \cite{Arutyunov:2004vx}.
Theoretically, one might also consider combinations 
of the various higher charges $q_s(x_j)$ of the original model. 
All of these generalizations are qualitatively different from the ones 
discussed in this paper because 
they explicitly refer to the positions $x_{j,k}$ of the 
Bethe roots. Here we will not discuss them further.

\subsection{A non-commutative gauge theory}

In the previous subsection we suggested that it is possible to deform
the spin chain of the $\mathcal{N}=4$  theory such that its
integrability is preserved while the Lagrangian of the deformed field
theory is not Lorentz invariant. These deformations appear to
fall in the class of deformations introduced in \secref{sec:DeformR} 
and  it is perhaps interesting to better understand
the structure of the deformed field theory leading to them.

It is certainly possible to extend the construction 
\cite{Lunin:2005jy} of the
supergravity duals of deformations of the type discussed in 
\secref{sec:ChTw} to include breaking of Lorentz and conformal invariance. 
Indeed, all one has to do is to identify the three commuting isometries 
of the AdS space and include them in the T-duality--shift--T-duality sequence 
of transformations. It is known from \cite{Hashimoto:1999ut,Maldacena:1999mh}
that such transformations performed along the isometries corresponding
to translations along the boundary of the AdS space in
the Poincar\'e patch correspond to the Moyal deformation of the boundary
theory. Our situation is different, however, since we are interested
in using the AdS isometries corresponding to the Cartan generators of
the four-dimensional superconformal group rather than the shift symmetries
manifest on the  Poincar\'e patch. Nevertheless, we expect to find
some kind of noncommutative field theory. 

Before discussing the deformation of the Lagrangian it is important to
decide the spacetime on which the theory is defined. In the case of
the $\mathcal{N}=4$ theory, conformal transformations map the theory
defined on $\IR^{4}$ and on $S^3\times\IR$ into each other. 
The deformation however breaks conformal symmetry and thus the 
deformation of the theory on the plane is in principle physically
different from the deformation of the theory defined on $S^3\times\IR$
and it is not \emph{a priori} clear which one corresponds to the
deformed spin chain introduced in the previous subsection. 

Starting from the properties of the potential gravity dual of the
deformed theory and borrowing from the experience with
the Lorentz-preserving deformations it is fair to guess that, in its
most general form, the deformation we are looking for replaces the
ordinary product of fields by
\[
XY~\mapsto~X*Y=e^{iC_{ab}h^a_X h^b_Y/2}\,X\,Y
\label{eq:lorentzbreak}
\]
where $h^a_{X,Y}$ are the Cartan generators of $\alg{psu}(2,2|4)$
acting on either $X$ or $Y$. 
The fact that the generators of the Cartan subalgebra commute 
among themselves and that they act following the Leibniz rule makes
this $*$-product somewhat similar with the Moyal product and thus
it is associative. 

In writing this expression we assumed that, from a string theory
perspective,  we are allowed to perform T-duality transformations
along any isometry, in particular that we can T-dualize the time
direction. This is usually problematic, since it yields the wrong 
sign for the kinetic terms of certain RR fields 
\cite{Cremmer:1998em,Cvetic:2002hi} and can be used only
as a solution-generating technique. It is therefore questionable
whether we should use it at all. This is similar to whether we should
use the charge vector ${\bf q}_{r_0}$ for twisting the gauge theory
spin chain.

In the case at hand a carefully-chosen deformation parameter -- such
that both the original and shifted coordinates
are time-like -- may render harmless the problem of the positivity of
the RR kinetic terms, so we will cautiously proceed along this
line.  
The gauge theory symmetry generator corresponding to global time
translations is a combination of the generator of scale transformations 
and a special conformal generator \cite{Horowitz:1998bj}. Since the
dilatation operator receives quantum corrections, it is not clear 
whether the corresponding $h$ appearing in \eqref{eq:lorentzbreak}
should be the quantum-corrected operator or only the classical part.
However, since the fundamental fields of the theory are not gauge
invariant, their anomalous dimensions exhibit gauge-fixing dependence
ambiguities. This suggests that, with the appropriate gauge-fixing, 
we may choose the tree-level generator of scale transformations
to appear in \eqref{eq:lorentzbreak}. This is analogous to the
observation in the previous section that ${\bf q}_{r_0}$ couples to
the classical limit of the Dynkin label $r$.

It is quite problematic to analyze in perturbation theory the 
consequences of the deformation \eqref{eq:lorentzbreak} of a field
theory on $\IR^4$. Perhaps the main obstacle is that the Cartan
generators depend explicitly on coordinates, $h\sim
x\partial_y-y\partial_x$ and thus the deformed Lagrangian is position
dependent.  Consequently, the momentum space calculations --
which are useful for ordinary noncommutative theories -- 
become inefficient for this kind of deformation. 

On $S^3\times \IR$ the situation is somewhat better because the
Lagrangian is not anymore position-dependent. Indeed, the Cartan
generators correspond to the two commuting isometries of 
$S^3\simeq \grp{SO}(4)/\grp{SO}(3)$ and translations along $\IR$. 
In this form they do not exhibit explicit position dependence 
and thus on $S^3\times \IR$  the deformation \eqref{eq:lorentzbreak} 
closely resembles the Moyal deformation. 
Explicit calculations are relatively difficult however
and perhaps the most efficient avenue involves dimensionally
reducing the theory to a quantum mechanical model with infinitely many
fields by expanding the four-dimensional fields in spherical harmonics
on $S^3$. Each such mode is clearly an eigenvector of the two Cartan
generators of $\grp{SO}(4)$ and, up to the cautionary words on
time-like T-duality and the identification of the generator dual to
global time translations in AdS space, of the third generator as well.
Thus, from the standpoint of this matrix quantum mechanics model, the
deformation \eqref{eq:lorentzbreak} appears identical (up to phases
involving the classical dimension of fields) with the one
involving the Cartan generators of the internal symmetry group. 
This is, of course,  not surprising since from the standpoint of the
$(0+1)$-dimensional quantum mechanical model, Lorentz and internal
symmetry  transformations are on equal footing. 

We will not write out explicitly the expression of the deformed field
theory Lagrangian. Clearly, it is quite complicated to directly compute
the dilatation operator and thus derive the spin chain
Hamiltonian from first principles.  It is however straight-forward to
use the deformed R-matrix to derive the planar dilatation operator at
one loop and it is probably possible to generalize it to include
non-planar interactions and thus generalize the calculation
\cite{Roiban:2004va} of correlation functions.%
Nevertheless, for the planar theory it is not necessary 
to have the Hamiltonian explicitly, but use the Bethe ansatz 
outlined in \eqref{sec:NCBethe} to determine the energy eigenvalues,
not only at one-loop, but probably at higher loops as well.
The benefit of this approach is that the Hamiltonian neither has to be 
computed nor evaluated. Both alternative steps would be 
extremely laborious when it comes to higher loops and long local
operators, see \cite{Beisert:2004ry}.

It is also not completely clear what the energy eigenvalues from
the Bethe ansatz correspond to in the noncommutative field theory.
They should be the generalization of anomalous dimensions of local
operators for a noncommutative theory. However, there are at least two
ways of constructing local composite operators in such a model.
One could either stick to the original definition, e.g.
\[\label{eq:LocalCommon}
\Op=\Tr \phi_{k_1}\phi_{k_2}\phi_{k_3}\ldots \phi_{k_L}+\ldots
\]
or use the $*$-product to concatenate the fields, e.g.
\[\label{eq:LocalStar}
\Op=\Tr \phi_{k_1}\ast\phi_{k_2}\ast\phi_{k_3}\ast\ldots \ast\phi_{k_L}+\ldots\,.
\]
The two pictures differ slightly although the energy eigenvalues should agree in the end
(see also \cite{Berenstein:2004ys}):
In the first, the action of the spin chain Hamiltonian is deformed uniformly. 
For the second, it is well-known that the spin chain Hamiltonian 
(i.e.~the planar interactions) is the same as for the commutative model.
This is at least almost true, but the interaction that wraps the trace
in \eqref{eq:LocalStar} is deformed, 
so here the Hamiltonian has a defect localized at the end of the trace. 
Similarly, the trace in \eqref{eq:LocalCommon} is manifestly cyclic, 
whereas the one in \eqref{eq:LocalStar} is not: 
the fields can be permuted cyclically, but not in any trivial way.

One can also cast the Bethe ansatz in both pictures.
A uniformly deformed Hamiltonian would give rise to twists whenever two
excitations cross their path. A term in the Bethe equation would
thus be written as
\[
\prod_{k'=1}^{K_{j'}}
\lrbrk{
e^{iA_{j,j'}}
\frac{u_{j,k}-u_{j',k'}+\ialf M_{j,j'}}{u_{j,k}-u_{j',k'}-\ialf M_{j,j'}}
}.
\]
When the deformation of the Hamiltonian is localized,
the phase shifts for crossing excitations are not modified. 
However, we have to consider the effect of 
excitations stepping across the defect. 
In the Bethe equations this would manifest as the term
\[
e^{iA_{j,j'}K_{j'}}
\prod_{k'=1}^{K_{j'}}
\frac{u_{j,k}-u_{j',k'}+\ialf M_{j,j'}}{u_{j,k}-u_{j',k'}-\ialf M_{j,j'}}\,.
\]
Clearly both pictures are equivalent and lead to the same final results.

\section{Discussion}
\label{sec:Concl}

In this article we have discussed a general method of twisting
integrable spin chains which is dual
to the twist described in \cite{Lunin:2005jy} on the supergravity side.
We have started out by describing the deformation of a generic
integrable quantum spin chain and its R-matrix. 
It is based on modifying the phase of the R-matrix 
depending on the Cartan charges of the involved sites.
We have shown that the deformed R-matrix still satisfies 
the Yang-Baxter equation and therefore our twist is an integrable deformation.
{}From the R-matrix for a spin chain with $\alg{u}(2|3)$ symmetry 
we have derived a set of twisted Bethe equations which generalizes
in a straight-forward fashion to arbitrary symmetry algebras.

The most general integrable twist is, however, not compatible
with the Feynman rules of twisted $\superN=4$ SYM. Conversely,
the most general deformation of the interactions of $\superN=4$ SYM
breaks integrability. We have derived the intersection of these two
requirements and obtained the deformation of the Bethe equations for
the complete ${\superN=4}$ spin chain model at one loop.
Excitingly, the deformation can be applied directly to the 
higher-loop Bethe ansatz proposed in \cite{Beisert:2005fw}
without corrupting its remarkable properties.
It is interesting to see that this depends crucially on 
the constraints from compatibility with Feynman rules;
it may be taken as an indication that the higher-loop Bethe ansatz indeed
respects the Feynman rules of the underlying field theory. 

Finally, we have generalized the twist to include deformations not only
of flavor symmetry, but also of spacetime symmetries. The resulting model 
is a sort of noncommutative field theory. On $\IR^4$ it differs
from the usual noncommutative theories in that the $*$-product employs
the Cartan generators of the Lorentz group rather than the momentum
generators. 
Alternatively, when using $S^3\times \IR$ as undeformed spacetime,
it becomes an ordinary noncommutative theory with noncommutativity 
involving two of the angles on $S^3$. 
Especially here, but also for the theory with twists restricted to flavor,
the Bethe equations are a valuable tool:
They allow to obtain planar anomalous dimensions without 
having to go through very laborious field theory computations,
which would be particularly cumbersome for the noncommutative theory.

It would be interesting to compare our results to the twisted analog
of the integrable structure of the classical superstring sigma model
on $AdS_5\times S^5$ \cite{Bena:2003wd}.  
For that one would have to repeat the construction
of the spectral curve and the finite gap method of
\cite{Kazakov:2004qf,Beisert:2005bm} for the twisted model 
proposed in \cite{Lunin:2005jy}. 
The resulting integral equations could then be 
compared to the thermodynamic limit of the
algebraic equations for the string chain 
\cite{Arutyunov:2004vx,Beisert:2004jw}
proposed in \cite{Beisert:2005fw}.

\subsection*{Acknowledgements}

We would like to thank J.~Maldacena for discussions. R.~R. would also
like to thank S.~Frolov and A.~Tseytlin for discussions and
collaboration on related topics. 

The work of N.~B.~is supported in part by the U.S.~National Science
Foundation Grant No.~PHY02-43680. Any opinions,
findings and conclusions or recommendations expressed in this
material are those of the authors and do not necessarily reflect the
views of the National Science Foundation.
The research of R.~R.~is supported in part by funds provided by the
U.S.D.O.E. under co-operative research agreement DE-FC02-91ER40671.

\bibliography{BetheTwist}
\bibliographystyle{nb}

\end{document}